\documentclass[11pt,draftclsnofoot,journal,onecolumn]{IEEEtran}

\IEEEoverridecommandlockouts

\usepackage[dvips]{graphicx}
\usepackage{subfigure}
\usepackage{amsmath}
\usepackage{amssymb}
\usepackage{verbatim}
\usepackage{url}
\usepackage{cite}
\usepackage{cases}

\renewcommand{\IEEEQED}{\IEEEQEDopen}

\begin{document}

\title{When Channel Bonding is Beneficial for Opportunistic Spectrum Access Networks}
\author{Shaunak Joshi, Przemys{\l}aw Pawe{\l}czak, Danijela \v{C}abri{\'c}, and John Villasenor%
\thanks{Copyright\copyright~2012 IEEE. Personal use of this material is permitted. However, permission to use this material for any other purposes must be obtained from the IEEE by sending a request to pubs-permissions@ieee.org.}
\thanks{Shaunak Joshi was with the Department of Electrical Engineering, University of California, Los Angeles. He is currently with Cisco Systems, Inc., San Jose, CA 95134, USA (email: shaunjos@cisco.com).}
\thanks{Przemys{\l}aw Pawe{\l}czak was with the Department of Electrical Engineering, University of California, Los Angeles. He is currently with Fraunhofer Institute for Telecommunications, Heinrich Hertz Institute, Einsteinufer 37, 10587 Berlin, Germany (email: przemyslaw.pawelczak@hhi.fraunhofer.de).}
\thanks{John Villasenor and Danijela \v{C}abri{\'c} are with the Department of Electrical Engineering, University of California, Los Angeles, 56-125B Engineering IV Building, Los Angeles, CA 90095-1594, USA (email: \{villa, danijela\}@ee.ucla.edu).}
\thanks{This work has been supported in part by the German Federal Ministry of Economics and Technology under grant 01ME11024 and by the German Federal Ministry of Education and Research under grant 01BU1224. Parts of this work has been published in the proceedings of IEEE GLOBECOM, 3--7 Dec., 2012, Anaheim, CA, USA~\cite{joshi_globecom_2012}.}}

\maketitle

\begin{abstract}
Transmission over multiple frequency bands combined into one logical channel speeds up data transfer for wireless networks. On the other hand, the allocation of multiple channels to a single user decreases the probability of finding a free logical channel for new connections, which may result in a network-wide throughput loss. While this relationship has been studied experimentally, especially in the WLAN configuration, little is known on how to analytically model such phenomena. With the advent of Opportunistic Spectrum Access (OSA) networks, it is even more important to understand the circumstances in which it is beneficial to bond channels occupied by primary users with dynamic duty cycle patterns. In this paper we propose an analytical framework which allows the investigation of the average channel throughput at the medium access control layer for OSA networks with channel bonding enabled. We show that channel bonding is generally beneficial, though the extent of the benefits depend on the features of the OSA network, including OSA network size and the total number of channels available for bonding. In addition, we show that performance benefits can be realized by adaptively changing the number of bonded channels depending on network conditions. Finally, we evaluate channel bonding considering physical layer constraints, i.e. throughput reduction compared to the theoretical throughput of a single virtual channel due to a transmission power limit for any bonding size.
\end{abstract}

\begin{IEEEkeywords}
Opportunistic spectrum access, medium access control, channel bonding, performance analysis.
\end{IEEEkeywords}

\section{Introduction}
\label{sec:introduction}

In wireless communications networks, bonding multiple frequency bands into one virtual channel improves the transmission speed for individual network users at the medium access control (MAC) layer, as the channel throughput is theoretically linearly dependent on channel bandwidth. Practical evaluation of channel bonding concepts has attracted substantial amount of attention, especially from the wireless local area networks (WLAN) research community~\cite{chandra_sigcomm_2008,arslan_conext_2010,tan_sigcomm_2010,moscibroda_icnp_2008}. Furthermore channel bonding is already present in some wireless networking standards, including IEEE 802.11n, IEEE 802.11ac and IEEE 802.22.

Intuitively, in networks utilizing channel bonding, each data flow will be transmitted faster as more channels are used for an individual connection. However, more occupied channels for one user translate into less data transmission opportunities for other network nodes~\cite{texas_wp_2003}. While this relationship is obvious, its analytical aspects have been relatively unexplored. Moreover, to the best of our knowledge, analytical models exploring this tradeoff and potential benefits of channel bonding MAC protocols do not exist. Surprisingly, the channel bonding concept has not been addressed analytically in the context of Opportunistic Spectrum Access (OSA) network behavior, specifically considering MAC layer features (particularly random access mode) and PU and SU traffic characteristics. As OSA networks utilize multiple, potentially non-contiguous channels, with primary users (PUs) randomly reappearing on their respective channels, it is important to understand the performance of channel bonding MAC protocols for OSA networks.

The objective of the paper is to understand the conditions under which it is beneficial to bond multiple frequency channels for OSA networks. To accomplish this, we present a mathematical framework based on a Markov chain analysis that enables the performance of channel bonding to be compared against classical OSA MAC protocols that do not use channel bonding. The framework enables the investigation of average MAC level network throughput as a function of the number of OSA network users, the number of frequency channels, channel bonding order, traffic level of secondary users (SUs), PU activity, and the spectrum sensing method, among many other parameters. The system model in our work is founded upon the model of~\cite{park_tmc_2011} and historically upon the model of~\cite{mo_tmc_2008}. This has provided for an ease of comparison of our results with earlier models of (non-bonded) multichannel MAC protocols. However, unique properties of our model (especially the allowance of channel bonding, which has not been considered in~\cite{mo_tmc_2008,park_tmc_2011}) resulted in a fundamentally new formulation of the Markov chain used in calculating the performance metrics.

In the literature related to channel bonding performance, one of the first studies of channel bonding is found in~\cite{chandra_sigcomm_2008}, where MAC protocols with channel bonding were advocated for throughput increase. It has been shown experimentally that channel bonding results in a large benefit for wireless networks. In this study only a small-scale network, i.e. with less than four nodes, was considered, thus it was unknown if the conclusions hold for large network sizes. In~\cite{arslan_conext_2010} the authors present measurement results of channel bonding functionality in IEEE 802.11n. They show through experiments that when transmitted power is fixed for WLANs with and without channel bonding, the network with bonded channels obtains smaller throughput than a network without this option enabled. To alleviate this problem authors propose an algorithm that adaptively merges the channels depending on the perceived link quality, where users with the same link quality are associated with the same access point. The algorithm has been evaluated only through experiments. In another study~\cite{tan_sigcomm_2010} a WLAN system with channel width adaptation was developed utilizing OFDM. A fixed number of channels were used, thus the effect of channel bonding adaptation on the network performance was not investigated. Summarizing, the investigation of channel bonding MAC protocols was strictly limited to the experimental platforms and it is unknown how channel bonding MAC protocols perform in scenarios not considered in the above mentioned studies.

A separate group of papers on channel bonding performance in an OSA context relates to early analytical studies. The benefit of channel bonding (in a broader context of frequency agility) has been evaluated analytically in~\cite{cao_dyspan_2010}. This study assumed that the OSA network assigns channels on a pre-detected pool of frequency bands, and MAC protocols are abstracted and not considered. In~\cite{yuan_mobihoc_2007} a spectrum allocation framework for OSA-based WLAN networks utilizing dynamically changing channel bonds was presented. First, a static and centralized channel allocation framework was proposed. Then a distributed MAC protocol utilizing a spectrum allocation scheme called B-Smart was discussed, and an analytical model based on a Markov chain analysis was used to assess the control channel throughput of B-Smart. The model used approximating expressions related to the probabilities of frame collision, frames being idle and successful frame transmission from~\cite{bianchi_jsac_2000}, which were applicable only to non-modified IEEE 802.11 systems with a distributed coordination function. The work in~\cite{yuan_mobihoc_2007} has been extended to WLAN networks in~\cite{moscibroda_icnp_2008}, where an analytical framework to evaluate efficiency of channel bonding for non-OSA networks with multiple access points has been proposed. However, as in the case of~\cite{yuan_mobihoc_2007}, the framework was applicable to static frequency allocation only, where the specifics of the protocol that network nodes use for communication were not a focus of the approach. The tool to model the problem was based on a linear programming formulation from which linear programming relaxation and a greedy heuristic were developed. A similar approach to analyze channel bonding was presented in~\cite{bansal_ohiotech_2010}, where analysis of an OSA ad-hoc network with static channel allocation for users cooperating with access points was analyzed. Non-contiguous channel allocation was allowed with more than one channel assigned to a single SU. In summary, all existing analytical tools to evaluate channel bonding are applicable to static frequency allocation and do not consider MAC protocol features. In the context of our work we need to mention~\cite{Park_arxiv_2010}, where a channel bonding feature for IEEE 802.22 networks was investigated. However, the proposed model only used approximations to reach a tractable solution, was applicable to static resource allocation networks based on OFDMA, and did not consider random access features of the MAC protocol.

Finally, we need to mention experimental OSA platforms with channel bonding. One of the first networks utilizing channel bonding for OSA network was the WhiteFi platform~\cite{bahl_sigcomm_2009}, i.e. a fully operational WLAN network working in the TV white spaces, which utilized varying size channels for data communications, i.e. 5, 10 and 20\,MHz. In WhiteFi channel bonding was possible only for adjacent channels. Performance of the platform was evaluated via simulations and experimentation with a limited set of prototype devices. A description of some of the algorithms and hardware used in WhiteFi platform was also discussed in~\cite{yuan_dyspan_2007}. Other OSA network prototypes utilizing channel bonding can be found in~\cite{yang_nsdi_2010}. The so called Jello framework fused multiple non-contiguous bands into one virtual channel, which was more flexible in this respect than WhiteFi. Indeed, the authors have shown a benefit of channel bonding, however measured only in terms of disruption rate and residual spectrum use. Implemented in a GNU Radio, Jello was evaluated in a testbed consisting of only a few nodes where the results were compared against simulation. In addition, off-line experimental studies of channel allocation for OSA, based on real-life measurement data of PU occupancy, have been presented in~\cite{kone_imc_2010}. Again, while this interesting study clearly showed how networks can benefit from accessing PU occupied channels, and how channel bonding benefits OSA networks, there are no analytical results that were presented therein.

The rest of the paper is organized as follows. The system model is introduced in Section~\ref{sec:system_model}, while the analytical model is presented in Section~\ref{sec:analytical_model}. Numerical results are presented in Section~\ref{sec:numerical_results} and the paper is concluded in Section~\ref{sec:conclusions}.

\section{System Model}
\label{sec:system_model}

We assume a single hop OSA ad-hoc network composed of $N$ nodes communicating with each other using multichannel MAC with a dedicated control channel (DCC), as discussed first in~\cite{park_tmc_2011,pawelczak_tvt_2009,mo_tmc_2008}\footnote{The reason for a single hop setup was due to the conformity and ease of comparison with the earlier studies on multichannel MAC protocols listed above. This makes it possible to easily compare analytical results on different OSA MAC protocol flavors with those available in this paper. In addition, many earlier studies on channel bonding in OSA networks, including prototype evaluation of channel bonding, considered single-hop domain as well, see e.g.~\cite{yuan_dyspan_2007,bahl_sigcomm_2009,kone_imc_2010}.}. Each node is assumed to transmit a saturated flow of framed data, where each new frame is generated with probability $p>0$. Parameter $p$, i.e. the connection request probability, governs the collision resolution strategy. A higher value of $p$ results in a higher number of collisions on the control channel, while a lower $p$ results in fewer collisions leading to a lower channel utilization. Furthermore, note that the assumption of traffic saturation for each node follows the classical assumption of many other works that have considered performance analysis of MAC protocols. As~\cite[Sec. III-B]{jha_twc_2011} notes ``(...) saturation throughput is a major performance measure to evaluate MAC protocols''. We refer for example to~\cite[Sec. II-C]{baldo_twc_2010}, \cite[Sec. IV]{zhang_jstsp_2011}, and \cite[Sec. 5]{lin_tvt_2012}, where such an assumption has been made while analyzing OSA MAC protocols. Note that by ``saturated'' traffic we mean traffic which is always sent by every node, however, at non-regular intervals governed by the control channel access probability $p$.

Nodes contend for the channel access by transmitting a Request to Send (RTS) frame via DCC, requesting a connection to other user of the network. The connection request is successful when only one SU requests a connection\footnote{Note that the application of our studies go well beyond single hop networks, as they are also applicable to centralized (access point-based) networks. The conversion of our analysis to centralized networks can be performed easily by adapting the construction of connection arrangement.}. The connection is established when the intended receiver responds with Clear to Send (CTS) frame on DCC when it is not involved in data exchange with other users. At the event of collision all connection requests are lost and users must contend for the DCC access after a random amount of time. This approach closely mimics the operation of the S-ALOHA protocol.

PU and SU transmission is slotted (SUs follow PUs' slot boundaries). The assumption on the slotted transmission of the SU and its synchrony with the PU is common in theoretical MAC analysis for OSA networks, see for example~\cite[Sec. II]{wang_twc_2012},~\cite[Fig. 2]{Liang_twc_2008},~\cite[Fig. 1]{Hoang_twc_2009},~\cite[Sec. III-A]{wangchun_jsac_2011}~\cite[Fig. 1]{cheng_jsac_2011},~\cite[Sec. II-A]{bian_jsac_2011}. Moreover, the assumption of PU/SU slot synchrony is due to conformity with our previous work on multichannel MAC protocols for OSA networks, i.e.~\cite{park_tmc_2011,pawelczak_tvt_2009}, where the same assumption has been made. This allows for an easy comparison of the results obtained in this paper with the previous results we have obtained on OSA multichannel MAC protocol performance. Furthermore, certain papers, e.g.~\cite{Papadimitratos_commag_2005,gronsund_pimrc_2009} do assume PU/SU synchrony in the practical network scenarios (in the case of~\cite{Papadimitratos_commag_2005}: secondary GSM network use, and in the context of~\cite{gronsund_pimrc_2009}: secondary IEEE 802.16 network use). Then, we emphasize that the PU/SU slot synchrony assumption allows for obtaining upper bounds on the throughput in comparison to slot-asynchronous interface, as it has been remarked in~\cite{gambini_twc_2008}. Please note however, that our model can indeed be generalized to a non-slotted PU/SU transmission. However, this would require further analysis on channel access policies, like those performed in~\cite{Gerihofer_commag_2007,Huang_infocom_2009}, which are beyond the scope of this paper.

Each slot is $T$ seconds long, with $T_s<T$ seconds of the spectrum sensing time performed at the beginning of the time slot by OSA network users and $T-T_s$ is equal to a length of one RTS/CTS frame exchange. Frame lengths are geometrically distributed such that the average frame length is $1/q(k)$ time slots, where $q(k)=\{C(T-T_s)/d\}k\beta(k)>1$ is the probability of a time slot being occupied by the frame transmission\footnote{In other words $q(k)$ is the probability that during frame transmission next time slot will be empty. This will end transmission and force a sender to contend over a control channel for a new virtual channel for new frame transmission.}, $C$ is the theoretical throughput of a single physical channel, $k$ is the number of physical channels bonded to form a virtual channel, $\beta(k)$ is the throughput reduction factor for a $k$-bonded virtual channel, where $\beta(1)=1$ and $\forall k>1$ $0\leq\beta(k)\leq1$, described in detail in Section~\ref{sec:phy_assumptions}, and $d$ is the size of the frame in bits. As we assume that the control channel throughput (in b/s) is the same as the throughput of data channel, i.e. $C$, we can easily calculate the length of the control packet as $d_p=\{C(T-T_s)\}/p$. In other words, in our model data and control packets, although of different lengths, are sent at the same rate.

\subsection{Multichannel Channel Bonding MAC Protocol Abstraction}
\label{sec:mac_abstraction}

In the present paper, the main difference with respect to the original design of the multichannel MAC with DCC (see again~\cite{park_tmc_2011,pawelczak_tvt_2009,mo_tmc_2008}) lies in assuming that each new connection will utilize $K$ out of $M=M_T-1$ channels, where $M_T$ is the total number of channels\footnote{In a network with a DCC, from the pool of all channels $M_T$, one channel is used for control data exchange only.} at time $t$, where $K$ is the maximum channel bond order. In the case of less than $K$ physical channels available for a new connection, all free channels are used for a frame transmission. We assume a set of accessible PU channels (which still can be randomly occupied by PUs) is known to the whole network and assigned unique identifiers. The sender, through the RTS frame, communicates to the receiver the information on the ordered set of available channels seen from its perspective, just like in~\cite[Sec. 4.1]{yuan_mobihoc_2007} (in the context of continuous frequency blocks). Next the receiver, through the CTS frame, replies with the set of ordered available channels from its perspective. After a successful RTS/CTS exchange both the sender and the receiver switch to the first $K$ channels that are common to both of them\footnote{In this paper we use RTS/CTS terminology to preserve the naming convention of~\cite{park_tmc_2011,mo_tmc_2008}. However, note that the purpose of the RTS/CTS transmission in the context of our work is different than for multi-hop networks. For example, in IEEE 802.11 networks (and their derivatives) RTS/CTS mechanism is used to resolve the hidden/exposed terminal problem~\cite{deng_tcom_2002}.}.

We assume that each communicating pair is able to bond non-adjacent frequency bands, with no guard bands considered, similar to~\cite{cao_dyspan_2010}. Thus, our framework serves as an upper bound on the performance of any channel bonding protocol considering physical layer constraints on the channel bonding process. Note that an analysis of the optimal channel bond resource allocation, considering guard bands, were considered recently, e.g., in~\cite{salameh_arizonatech_2011}. In the remainder of this paper we refer to this channel bonding MAC protocol as flexible channel bonding.

\subsection{Primary User and Detection Process}
\label{sec:primary_user_detection}

Each PU channel of bandwidth $W$ at time slot $t$ is randomly occupied by the PU with probability $q_p$, where $q_p$ is geometrically distributed. While not all PU traffic is memoryless, refer e.g. to studies of~\cite{geirhofer_commag_2007} that prove such a statement, many studies reveal that the complete opposite is true. Various PU systems can indeed be correctly described by a memoryless process. For example, in~\cite{wellens_phycom_2009}, one of the most complete long-term studies of channel occupancy by various wireless systems, the geometric distribution is quite common, constituting almost 60\% of PU traffic distributions for systems of interest, including DECT, GSM (900/1800) and 2.4\,GHz ISM\footnote{Note that the studies of non-memoryless distributions' impact on multichannel MAC protocols were considered in our earlier studies. We refer to~\cite[Sec. 5.2.6]{park_tmc_2011} for further discussion.}.

Due to a single hop domain analysis we can reasonably assume for tractability that each OSA node performs detection individually and the decision on the PU state is the same for each node in the OSA network. Furthermore, each OSA node is equipped with a wideband spectrum sensor to obtain information about all PU channels during the sensing phase\footnote{Theoretically, channel bonding would be possible with a narrowband/single antenna sensor, but only under certain limiting conditions. That is, slow PU activity with fast channel sensing time, i.e. orders of magnitude faster than the frame sending rate of the SUs. This would potentially allow to sense all PU channels in a round robin fashion by the SU.}. Decision on the PU state is prone to errors due to false alarm, $p_f$, or mis-detection, $1-p_{d}$. Assuming Rayleigh fading and Gaussian noise the respective expressions are given in~\cite[Eq. (1)]{pawelczak_tvt_2009} and~\cite[Eq. (3)]{pawelczak_tvt_2009}. For those expressions we will give respective values for the PU channel bandwidth $W$ and SNR $\bar\gamma$ of the PU signal at the SU energy detector with detection threshold $\Omega$, resulting in required $p_f$ and $p_d$, when presenting the numerical results. We assume that the OSA network loses throughput only due to false alarms, and it can successfully deconstruct a frame on the arrival of a PU during a mis-detection. On the other hand, we assume that the detection of a PU on any of the physical channels constituting one virtual channel causes termination of an entire frame transmission. Although this is a very conservative strategy it simplifies the analysis significantly, while also serving as a lower bound on the obtained MAC throughput considering frame disruption resolution strategies.

\subsection{Physical Layer Assumptions}
\label{sec:phy_assumptions}

In the analysis we consider two cases with respect to virtual channel throughput. In the first case we assume that throughput of a virtual channel with $k$ bonds is exactly $kC$, as in~\cite{cao_dyspan_2010}, i.e. $\forall k$ $\beta(k)=1$. In the second case we assume, due to the fixed power that users can emit per virtual channel, the throughput for a $k$-bonded channel is $\beta(k)kC$, where $\forall k>1$ $\beta(k)<1$ and $\beta(1)=1$. Note that the function $\beta(k)$ can be defined arbitrarily $\forall k>1$, depending on what physical layer constraint is considered, e.g., it can mimic the exponential decay of throughput for virtual channels of increasing orders, or it can mimic step-wise throughput reduction with increasing $k$.

\subsection{Comparison with other Multichannel MAC Protocols}
\label{sec:comparison_protocols}

For comparison in this study we also consider other multichannel MAC protocols. The first protocol is a classic multichannel MAC protocol analyzed first in~\cite{pawelczak_tvt_2009}, where bonding is not allowed, i.e. channel bonding protocol with $K=1$. The second protocol, denoted as $K$-only bonding multichannel MAC, is a less flexible version of the main channel bonding MAC protocol. That is, when a newly admitted connection sees fewer than $K$ free channels, then this connection is blocked and the user contends again for the channel access. This protocol mimics the operation of networks which are able to bond only a fixed number of channels per newly arriving frame. The third protocol is an adaptive version of the considered channel bonding multichannel MAC protocol, that changes the order of channel bonding dynamically depending on the network and/or traffic conditions. Details of the protocol will be presented in Section~\ref{sec:results_adaptive_bonding}. In the paper we will study the flexible channel bonding MAC protocol via analysis, while $K$-only channel bonding and adaptive channel bonding will be considered via simulations.

\section{Analytical Model}
\label{sec:analytical_model}

To calculate network-wide average MAC layer network throughput we propose to use a Markov chain analysis framework\footnote{Use of the Markov chain as an analytical tool to understand the operation of MAC protocols is a standard approach in communications literature, e.g. see recent work of~\cite{wang_jsac_2011,wang_tvt_2010} analyzing (non-channel bonding) MAC protocols for OSA. Therefore, we follow the same approach here.}. The model presented here extends the work of~\cite{park_tmc_2011} to the case of channel bonding. Therefore we follow the same naming convention and use the same definition of connection arrangement and termination probabilities (as the system model in this work follows those of~\cite{park_tmc_2011} and the earlier model of~\cite{mo_tmc_2008}). However, the structure of the Markov chain, formulation of transition probabilities, introduction of a new function: connection preemption probability, and calculation of performance metrics are fundamentally new and unique to our work.

The roadmap of this section is as follows. First, in Section~\ref{sec:chain_definition} we introduce the definition of a Markov chain used to analyze our proposed MAC protocol. Next, in Section~\ref{sec:arrangement_probability} and Section~\ref{sec:termination_probability} we present the definition of the connection arrangement and connection termination probability, respectively, while in Section~\ref{sec:overlap_probability} we introduce the definition of the connection preemption probability. These three functions govern the way connections are established and terminated within the OSA network of interest. Section~\ref{sec:transition_probability} derives the complete Markov chain transition probabilities (using the three above mentioned functions) which are later used in Section~\ref{sec:metric_calculation} to derive the performance metrics of interest.

\subsection{Definition of a Markov Chain}
\label{sec:chain_definition}

Assuming without loss of generality that $N>M$, i.e. that the OSA network needs to compete for limited resources, let $\mathbf{x}_t=(x_{1,t},\ldots,x_{K,t})$ denote a state of the Markov chain at time $t$, where state, $0\leq x_{i,t} \leq \lfloor M/i \rfloor$ denotes the number of active connections occupying $i$ physical channels at time $t$. Then, $\mathcal{X}_t={\mathbf{x}_{t}^{(1)},\ldots,\mathbf{x}_{t}^{(y)}}$ is the set of all possible states for the considered system at time $t$, and $y$ is the size of $\mathcal{X}_t$. Furthermore, let
\begin{align}
r_{t}^{(t+1)}=\Pr(x_{1,t+1}=&a_{1},\ldots,x_{K,t+1}=a_{K}|\nonumber\\&\qquad x_{1,t}=b_{1},\ldots,x_{K,t}=b_{K}).
\label{eq:transition}
\end{align}
Then the steady state probability vector $\pi_{x_1,\ldots,x_K}$ is obtained from transition probabilities as
\begin{equation}
\pi_{x_1,\ldots,x_K}=\lim_{t\rightarrow\infty}\Pr(x_{1,t}=x_{1},\ldots,x_{K,t}=x_{K}),
\label{eq:steady_state_derivation}
\end{equation}
while $\mathcal{X}=\lim_{t\rightarrow\infty}\mathcal{X}_t$

\subsection{Connection Arrangement Probability}
\label{sec:arrangement_probability}

The probability that $j$ new connections at time $t+1$ are successfully admitted through a control channel while $m$ pairs of users actively exchange data on the virtual channel at time $t$ is defined as~\cite[Eq. (15)]{park_tmc_2011}
\begin{equation}
S_{m}^{(j)}=\begin{cases}
(N-2m)p(1-p)^{N-2m-j}, & j=1,\\
1-S_{m}^{(1)}, & j\neq1,
\end{cases}
\label{eq:arrangement}
\end{equation}
where, without loss of generality, the common control channel is assumed to be absent from the PU activity. If one wants to assume that DCC is also randomly occupied by the PU, then (\ref{eq:arrangement}) can be easily replaced with~\cite[Eq. (15) and (16)]{park_tmc_2011}\footnote{Note that the application of our model to ad ahoc networks allows for the analysis of infrastructure-based networks as well. For example, in the simplest case one needs to change the definition of $S_{m}^{(j)}$ to $(N-m)p(1-p)^{N-m-j}$ for $j=1$. This would mean that every new connection request involves only a single user (instead of two), e.g. a single user requesting connection to an access point. Obviously, other definitions of $S_{m}^{(j)}$ can be used in our work that would reflect the operation of infrastructure-based networks better. For example, one can use~\cite[Eq. (3)]{zhou_ieeetwc_2008}, i.e. definition of a PRACH-like control channel used in 3GPP-based networks.}.

\subsection{Connection Termination Probability}
\label{sec:termination_probability}

The probability that $j$ connections at time $t+1$ finish transmission out of $m$ connections using $k$-bonded virtual channels at time $t$, is defined as~\cite[Eq. (14)]{park_tmc_2011}
\begin{equation}
^{(k)}T_{m}^{(j)}=\begin{cases}
\binom{m}{j}q(k)^{j}(1-q(k))^{m-j}, & m \geq j > 0,\\
0, & \text{otherwise}.
\end{cases}
\label{eq:termination}
\end{equation}

\subsection{Connection Preemption Probability}
\label{sec:overlap_probability}

The probability that existing SU connections are preempted by PUs is defined as
\begin{align}
P_{\mathbf{x},z}^{(f,\mathbf{r})}=\sum_{i=1}^{Y}&\left[\sum_{j=1}^{K}\sum_{m=1}^{c_{i,j}}\sum_{n=1}^{a_{j}}\binom{j}{l_{m,n}^{\left(i,j\right)}}+\sum_{m=1}^{c_{K+1}}\sum_{n=1}^{f}\binom{1}{l_{m,n}^{\left(i,K+1\right)}}\right]\nonumber\\&\qquad\times q_{c}^{z}\left(1-q_{c}\right)^{M-z},
\label{eq:pxyt}
\end{align}
where $q_{c}=q_{p}p_d+(1-q_{p})p_{f}$~\cite[Eq. (13)]{park_tmc_2011} is the state of the PU observed by the SU network\footnote{Note again that $p_f$ results from a certain energy detection process for a required $p_d$; see Section~\ref{sec:primary_user_detection} for details.}, $z$ is the total number of PU occupancies on physical channels at time $t$, $f=M-\sum_{i=1}^{K}ib_{i}$, with $b_i$ given in (\ref{eq:transition}), is the number of PU occupancies on physical channels not used by SU pairs, $\mathbf{r}_{t+1}=[r_{1,t+1},\ldots,r_{K,t+1}]$, where $r_{k,t+1}$ is the total number of preemptions for connections using $k$-bonded virtual channels at time $t+1$, and $Y$ is the total number of combinations of PU occupancy across all physical channels that account for $\mathbf{r}_{t+1}$ SU preemptions. Following (\ref{eq:arrangement}) and (\ref{eq:termination}), to simplify the notation in (\ref{eq:pxyt}) we have omitted time index $t$ such that $\mathbf{x}=\mathbf{x}_t$ and $\mathbf{r}=\mathbf{r}_{t+1}$. To define $l_{m,n}^{(i,j)}$ we introduce
\begin{equation}
\mathbf{\Lambda}=\left[\begin{array}{cccc}
\mathbf{\lambda_{1,1}} & \cdots & \mathbf{\lambda_{1,K}} & \mathbf{\lambda_{1,K+1}}\\
\vdots & \ddots & \vdots & \vdots\\
\mathbf{\lambda_{Y,1}} & \cdots & \mathbf{\lambda_{Y,K}} & \mathbf{\lambda_{Y,K+1}}
\end{array}\right],
\label{eq:Lambda}
\end{equation}
where $\forall j: 1\leq j\leq K$ $\mathbf{\lambda_{i,j}}$ is a matrix, the rows of which represent the distribution of PUs that can occupy channels to ensure $r_i$ SU preemptions, and $\lambda_{i,K+1}$ is a matrix, the rows of which contain the distribution of PUs on $f$ unoccupied channels, while still maintaining $r_i$ preemptions. Specifically, $\forall j: 1\leq j\leq K+1$,
\begin{equation}
\mathbf{\mathbf{\lambda_{i,j}}}=\left[\begin{array}{cccc}
l_{1,1}^{\left(i,j\right)} & \cdots & l_{1,a_{K+1}}^{\left(i,j\right)}\\
\vdots & \ddots & \vdots\\
l_{c_{j},1}^{\left(i,j\right)} & \cdots & l_{c_{j},a_{K+1}}^{\left(i,j\right)}
\end{array}\right],
\end{equation}
where $l^{\left(i,j\right)}_{m,n}$ indicates the number of PU occupancies on one of $a_j$ currently active $j$-bonded connections in order to ensure a total of $r_i$ SU preemptions across all such active connections. Note that $a_{K+1}$ is the current number of free channels and is not related to the number of bonded connections, $a_j$, defined $\forall j: 1\leq j\leq K$. The term $c_{i,j}$ represents the total number of combinations of arranging PUs on $a_j$ currently active $j$-bonded connections to ensure $\mathbf{r}$ preemptions for the $i$th row of $\mathbf{\Lambda}$.

\subsection{Transition Probabilities}
\label{sec:transition_probability}

First, we introduce the indicator function: $I(\eta)=1$ if $\eta$ is true and $I(\eta)=0$ otherwise, where $\eta$ represents the logical condition under test. We further introduce supporting notation that is used in the derivation of the transition probabilities: $\mathbf{0}_K$ denotes the vector of length $K$ containing only zeros; $\mathbf{{\delta}}_i$ denotes a vector of length $K$ containing zeros except with a 1 value at index $i$; $A=\sum_{i=1}^{K}a_{i}$, which is the sum of all the current connections; and $\alpha=\min\left(f,K\right)$. The transition probabilities are grouped into four broad cases shown below.

\subsubsection{If $\exists!i:b_{i}-a_{i}=1$ and $\forall j\ne i:b_{j}-a_{j}=0$}
\label{sec:case_1}

This case deals with the event in which there is only one new connection being made while there are no terminating connections. Then
\begin{equation}
r_{t}^{(t+1)}=\prod_{i=1}^{K}{}^{(i)}T_{a_i}^{(0)} S_{A}^{(1)}P_{\mathbf{x},z}^{(f,\mathbf{0}_K)}.
\label{eq:rt1c1}
\end{equation}
It is noted that the event is reflected directly in the right superscripts of the $^{(x)}T_{y}^{(z)}$ functions while there is exactly one connection generation seen in the superscript of $S_{x}^{(y)}$. As there are no terminations due to preemption by PUs, hence the use of all the all-zeros vector $\mathbf{0}_K$.

\subsubsection{If $\forall i:b_{i}-a_{i}=0$}
\label{sec:case_2}

In the second case we explore the event in which there is no change in the number of connections within a time slot. There are two subcases as listed below.

First, if $f>0$, deals with the condition in which the next state is not a state in which connections occupy all available channels. The state in which all channels are occupied is referred to as an \textit{edge} state. Then
\begin{equation}
r_{t}^{(t+1)}=\sum_{i=1}^{3}{}^{(i)}r_{t}^{(t+1)},
\label{eq:rt}
\end{equation}
where $^{(1)}r_{t}^{(t+1)}$ is given as (\ref{eq:rt1c1}) replacing $S_{A}^{(1)}$ with $S_{A}^{(0)}$, i.e.
\begin{equation}
^{(1)}r_{t}^{(t+1)}=\prod_{i=1}^{K}{}^{(i)}T_{a_{i}}^{(0)} S_{A}^{(0)}P_{\mathbf{x},z}^{(f,\mathbf{0}_K)},
\label{eq:2a1}
\end{equation}
\begin{equation}
^{(2)}r_{t}^{(t+1)}=
\begin{cases}
\begin{split}&\prod_{i=1}^{K-1}{} ^{(i)}T_{a_i}^{(0)}{}^{(K)}T_{a_{K}}^{(1)}\\&\quad\times S_{A}^{(1)}P_{\mathbf{x},z}^{(f,\mathbf{0}_K)},\end{split} & \begin{split}&\text{if } f \geq K, \\&z=0, \\&a_{K}>0,\end{split}\\
0, & \text{otherwise,}
\end{cases}
\label{eq:2a2}
\end{equation}
\begin{equation}
^{\left(3\right)}r_{t}^{(t+1)}= \begin{cases}
\begin{split}&\prod_{i=1}^{K}{} ^{(i)}T_{a_i}^{(0)}{}S_{A}^{(1)}\\&\quad\times P_{\mathbf{x}+\mathbf{\delta_{\alpha}},z}^{(f-\alpha,\mathbf{0}_K+\mathbf{\delta_{\alpha}})},\end{split} & z>0,\\
0, & \text{otherwise.}
\end{cases}
\label{eq:2a3}
\end{equation}

Equation~(\ref{eq:2a1}) corresponds to the event in which there are no terminating connections and no preemptions due to PUs. Equations~(\ref{eq:2a2}) and (\ref{eq:2a3}) are conditional and are dependent on whether PUs are absent or present, respectively, in order to preempt SUs. In~(\ref{eq:2a2}) a SU bonding $K$ channels terminates a connection while another SU generates a frame ensuring the same connection state. In~(\ref{eq:2a3}) an additional SU with $K$-bonded virtual channel makes a connection, but is preempted by a PU. Note that the current connection state and the number of free channels is adjusted by a factor of $\alpha$, which is the number of channels bonded where an additional connection is made.

Second, if $f=0$, represents the transition necessary when the next state is an edge state. Then $r_{t}^{(t+1)}$ is defined as (\ref{eq:rt}) with the respective probabilities defined as
\begin{equation}
^{(1)}r_{t}^{(t+1)}\!=\!\prod_{i=1}^{K} {}^{(i)}T_{a_{i}}^{(0)}\!\! \left(S_{A}^{(0)}P_{x,z}^{(f,\mathbf{0}_K)}\!+\!S_{A}^{(1)}P_{\mathbf{x},z}^{(f-\alpha,\mathbf{0}_K)}\right),
\label{eq:2b1}
\end{equation}
\begin{equation}
^{\left(2\right)}r_{t}^{(t+1)}= \begin{cases}
\begin{split}&\prod_{i=1}^{K-1}{} ^{(i)}T_{a_i}^{(0)}{}^{(K)}T_{a_{K}}^{(1)}\\ &\quad\times S_{A}^{(1)} P_{\mathbf{x},z}^{(f,\mathbf{0}_K)},\end{split} & \begin{split}&\text{if }\sum_{i=1}^{K}ib_{i}<M,\\& M\bmod{K}=0,\\& a_{K}>0,\end{split}\\
0, & \text{otherwise,}
\end{cases}
\label{eq:2b2}
\end{equation}
and $^{(3)}r_{t}^{(t+1)}$ defined in the same way as (\ref{eq:2a3}). Subcase $^{(1)}r_{t}^{(t+1)}$ (defined as (\ref{eq:2b1})) represents the event in which there is no additional connections or terminations ensuring the same connection state, and also contains an excess condition where one additional connection is blocked because of all channels being occupied, i.e. the edge case. Subcase $^{\left(2\right)}r_{t}^{(t+1)}$ (defined as (\ref{eq:2b2})) and subcase $^{\left(3\right)}r_{t}^{(t+1)}$ (defined in the same way as (\ref{eq:2a3})) are conditional and represent the case in which an extra connection can be preempted by an SU and an additional connection with $K$ bonded channels replaces a terminating connection, respectively.

\subsubsection{If $\exists i:b_{i}-a_{i}<0$ and $z>0$}
\label{sec:case_3}

This case represents the event in which there is at least one terminating connection and preemptions are also possible because of the presence of PUs.

Before we describe this case in detail, we introduce further notation. Let $\mathbf{\Theta}$ be a matrix where each row contains the combination of allowable terminations for the particular state transition, where the total number of such combinations equals $c$, i.e.,
\begin{equation}
\mathbf{\Theta}=\left[\begin{array}{cccc}
\theta_{1,1} & \theta_{1,2} & \cdots & \theta_{1,K}\\
\vdots & \vdots & \ddots & \vdots \\
\theta_{c,1} & \theta_{c,2} & \cdots & \theta_{c,K}
\end{array}\right],
\label{eq:Theta}
\end{equation}
where the rows of $\Theta$ are denoted as $[\mathbf{\theta}_{1},\ldots,\mathbf{\theta}_{K}]^{T}$. Furthermore, let $\mathbf{\tau}=\mathbf{x}_{t+1}-\mathbf{x}_{t}$, where $\tau_j$ is the $j$th element of $\mathbf{\tau}$. We consider the following three subcases.

First, if $f=0$ and $\forall i,j:\tau_j-\theta_{i,j}>0$, there are terminations possible due to overlap, i.e. $\forall j:\tau_j-\theta_{i,j}>0$ including terminations according to the rows of $\mathbf{\Theta}$ with both none and an additional excess connection generation, i.e.
\begin{align}
r_{t}^{(t+1)}=\sum_{i=1}^{c}\prod_{j=1}^{K}{}^{(j)}T_{a_{j}}^{(\theta_{i,j})}\left(S_{A}^{(0)}+S_{A}^{(1)}\right) P_{\mathbf{x}-\mathbf{\theta}_{i},z}^{(0,\mathbf{\tau}-\mathbf{\theta}_{i})}.
\end{align}

Second, if $f>0$ and $\forall j:\tau_j-\theta_{i,j}>0$ and $\nexists i:b_{i}-a_{i}=1$, there are terminations possible due to overlap and although free channels are available, an increase in connection does not occur. That is
\begin{align}
r_{t}^{(t+1)}&=\sum_{i=1}^{c}\left[\prod_{j=1}^{K}{}^{(j)}T_{a_{j}}^{(\theta_{i,j})}S_{A}^{(0)} P_{\mathbf{x}-\mathbf{\theta}_{i},z}^{(f+\sum_{j=1}^{K}j\theta_{i,j},\tau-\theta_{i})}\nonumber\right.\\&\left.+I\left(\sum_{j=1}^{K}\left(a_{j}-\theta_{i,j}\right)+1-z\leq\sum_{j=1}^{K}jb_{j}\right)\left\{\prod_{j=1}^{K}{}^{(j)}T_{a_{j}}^{(\theta_{i,j})}\right.\right.\nonumber\\&\left.\left.\times S_{A}^{(1)}P_{\mathbf{x}+\mathbf{\delta_{\alpha}}-\theta_{i},z}^{(f+\sum_{j=1}^{K}j\theta_{i,j}-\alpha,\mathbf{\tau}-\mathbf{\theta}_{i})}+I\left(x_{\alpha}-\theta_{i,\alpha}=b_{\alpha}\right)\right.\right.\nonumber\\&\left.\left.\times{}^{(1)}T_{a_{1}}^{(\theta_{i,1})}\cdots^{(\alpha)}T_{a_{\alpha}}^{(\theta_{i,\alpha}+\mathbf{\delta_{\alpha}})}\cdots{}^{(K)}T_{a_{K}}^{(\theta_{i,K})} \right.\right.\nonumber\\&\left.\left.\times S_{A}^{(1)} P_{\mathbf{x}-\mathbf{\theta}_{i},z}^{(f+\sum_{j=1}^{K}j\theta_{i,j},\mathbf{\tau}-\mathbf{\theta}_{i})}\right\} \right].
\label{eq:3b}
\end{align}
In (\ref{eq:3b}) the logical argument in the outermost indicator function is used to determine if there are enough PUs present for preemption in the event that an additional connection generation occurs and terminations are insufficient. The logical argument in the innermost indicator function is used to determine if terminations for $\alpha$-bonded connections alone are sufficient to reach the desired end state.

And finally, if $f>0$ and $\exists !i:b_{i}-a_{i}=1$, there are free channels and there exists one and only one additional connection generation the following transition probability is used, i.e.
\begin{align}
r_{t}^{(t+1)}&=\sum_{i=1}^{c}\left[I\left(\sum_{j=1}^{K}\left(a_{j}-\theta_{i,j}\right)+1-z\leq\sum_{j=1}^{K}jb_{j}\right)\nonumber\right.\\&\left.\qquad\times\prod_{j=1}^{K}{}^{(j)}T_{a_{j}}^{(\theta_{i,j})} S_{A}^{(1)}P_{\mathbf{x}-\mathbf{\theta}_{i},z}^{(f+\sum_{j=1}^{K}j\theta_{i,j},\mathbf{\tau}-\mathbf{\theta}_{i})}\right].
\end{align}
Note that as in~(\ref{eq:3b}) the indicator function is used to determine if preemptions combined with terminations are sufficient to reach the desired end state.

\subsubsection{If $\exists i:b_{i}-a_{i}<0$ and $z=0$}
\label{sec:case_4}

Similar to case described in Section~\ref{sec:case_3}, this case covers the event in which a termination occurs, however there are no preemptions possible because there are no PUs occupying the system in the end state. That is
\begin{equation}
r_{t}^{(t+1)}={}^{(1)}r_{t}^{(t+1)}+{}^{(2)}r_{t}^{(t+1)}.
\end{equation}
For the purpose of simplifying the notation we introduce the following supporting function
\begin{align}
\xi\left(g\right)&=\sum_{i=1}^{c}\left[I\left(\sum_{j=1}^{K}\theta_{i,j}>\sum_{j=1}^{K}\tau_{j}\right)\prod_{j=1}^{K}{}^{(j)}T_{a_{j}}^{(\theta_{i,j})}\nonumber\right.\\&\left.\qquad\times P_{\mathbf{x}-\theta_{i},0}^{(f+\sum_{j=1}^{K}j\theta_{i,j},\mathbf{0}_K)}g\right].
\end{align}
Then
\begin{subnumcases}{\!\!\!\!\!\!^{(1)}r_{t}^{(t+1)}\!\!=\!\!}
\label{eq:4_1}
\xi\left(S_{A}^{(0)}\right), \!\!\! & $\begin{split}\text{if }f>0,\hat{a}_j=\hat{b}_j,\end{split}$\label{eq:4_1a}\\
\xi\left(S_{A}^{(0)}+S_{A}^{(1)}\right), \!\!\! & $\begin{split}\text{if } f=0, \hat{a}_j=\hat{b}_j,\end{split}$\label{eq:4_1b}\\
\xi\left(S_{A}^{(1)}\right), \!\!\! & $\begin{split}\text{if }\hat{a}_j\neq\hat{b}_j,\end{split}$\label{eq:4_1c}\\
0, & \text{otherwise}\label{eq:4_1d},
\end{subnumcases}
where $\hat{a}_j=\sum_{j=1}^{K}(a_{i}-\theta_{i,j})$ and $\hat{b}_j=\sum_{j=1}^{K}b_{j}$. The first two conditions, i.e.~(\ref{eq:4_1a}) and~(\ref{eq:4_1b}), represent the non-edge and edge cases, respectively, for when there are no additional generations in the system, i.e. the terminations subtracted from the current state is equal to the end state. The third condition, i.e.~(\ref{eq:4_1c}), represents the case in which there is an additional generation accompanying the one or more terminations in the current connection state.

Furthermore
\begin{equation}
\! ^{(2)}r_{t}^{(t+1)}=
\begin{cases}
\begin{split}&{}^{(1)}T_{a_{1}}^{(\theta_{i,1})}\cdots\\&\quad\times^{(\alpha)}T_{a_{\alpha}}^{(\theta_{i,\alpha}+\mathbf{\delta_{\alpha}})}\cdots\\&\quad\times{}^{(K)}T_{a_{K}}^{(\theta_{i,K})} S_{A}^{(1)}\\&\quad\times P_{\mathbf{x}-\mathbf{\theta}_{i},0}^{(M-\sum_{j=1}^{K}jb_{j},\mathbf{0}_K)},\end{split} & \!\!\!\!\! \begin{split}&\text{if }f>0,\\& a_{\alpha}>\tau_{\alpha},\\& \nexists i:b_{i}-a_{i}=1,\end{split}\\
0, & \!\!\!\!\! \text{otherwise.}
\end{cases}
\label{eq:4_2}
\end{equation}
In (\ref{eq:4_2}) there are additional free channels, but there is no increase in any of the bonded channels. In this case an additional connection generation in the $\alpha$-bonded channel is terminated by adding a $\mathbf{\delta_{\alpha}}$ to the $\alpha$th term of the termination vector row $\theta_i$.

\subsection{Performance Metric Calculation}
\label{sec:metric_calculation}

Given the steady state matrix $\pi_{x_1,\ldots,x_K}$ we can compute the performance metrics of the system of interest. The system throughput is calculated as
\begin{equation}
R=C\frac{T-T_s}{T}\sum_{i=1}^{y}\sum_{j=1}^{K}j\mathbf{x}^{(i)}(j)\beta(j)\pi_{x_1,\ldots,x_K}(i).
\label{eq:throughput}
\end{equation}
Note that the channel utilization can be calculated as
\begin{equation}
U=\frac{T-T_s}{T}\frac{\sum_{i\in \mathcal{X}}\sum_{j=1}^{K}\mathbf{x}_{i,j}j\pi(i)}{M}.
\label{eq:utilization}
\end{equation}
Also note, that in (\ref{eq:throughput}) and (\ref{eq:utilization}) both metrics are multiplied by the factor $(T-T_s)/T$, due to incurred spectrum sensing overhead.

To conclude, in the context of our model the metrics of (\ref{eq:throughput}) and (\ref{eq:utilization}) are the fundamental descriptors of the channel bonding MAC performance in OSA. In the next section we will present numerical results to gain insight on the operation of such a protocol based on these two metrics.

\section{Numerical Results}
\label{sec:numerical_results}

We present a set of numerical results to evaluate channel bonding for OSA from a MAC layer perspective. First, in Section~\ref{sec:result_pu_impact}, we assess the impact of PU activity level $q_p$, introduced in Section~\ref{sec:primary_user_detection}, on the system throughput. Secondly, in Section~\ref{sec:resutls_ch_pool}, we investigate the effect of the ratio of channel pool size to the total number of OSA network users on the system throughput. In Section~\ref{sec:results_virtual_throughput}, we investigate the effect of individual virtual channel throughput, $kC\beta(k)$, introduced in Section~\ref{sec:phy_assumptions}, on total system throughput. Then, in Section~\ref{sec:results_pu_distruption} we consider the impact of a virtual channel disruption strategy on system throughput, followed by studies on adaptive channel access control in Section~\ref{sec:results_fairness}. In Section~\ref{sec:results_adaptive_bonding} we discuss the construction of an adaptive channel bonding MAC protocol. Finally, in Section~\ref{sec:non_geometric} we present results on the effect of non-geometric PU duty cycle distribution on the performance of channel bonding. In the subsequent sections we focus on the throughput metric $R$ solely, since channel utilization $U$ is directly proportional to $R$. Since the issues of random access phase and the effect of spectrum sensing layer were investigated in earlier studies of multichannel MAC protocols~\cite{park_tmc_2011,pawelczak_tvt_2009}, we do not explore them here.

For the sake of clarity and without loss of generality, we present the results by grouping them into two network scenarios unless otherwise stated: (a) PU channel pool $M=4$, number of users $N=12$, and frame size $d=5$\,kB, representing a small-scale network; and (b) $M=12$, $N=40$, $d=20$\,kB, representing a large-scale network. The remaining parameters common to both scenarios are (unless otherwise stated): control channel access probability $p=e^{-1}/N$ (as in~\cite{park_tmc_2011,pawelczak_tvt_2009}), physical channel throughput $C=200$\,kB, slot length $T=1$\,ms, sensing time $T_s=0.1$\,ms, throughput reduction function $\forall i$ $\beta(i)=1$ (perfect bonding scheme), $\bar\gamma=0$\,dB PU signal received power at each SU detector, detection threshold $\Omega=17.8$\,dB and $W=200$\,kHz bandwidth for which $p_d=0.9$ and $p_f=0.02$. We have tested the performance of channel bonding schemes for realistic bonding order, i.e. $K\in\{2,3\}$, as investigated in, e.g.,~\cite{arslan_conext_2010,tan_sigcomm_2010,chandra_sigcomm_2008}.

\subsection{Impact of PU Activity on System Throughput}
\label{sec:result_pu_impact}

\begin{figure}
\centering
\subfigure[$M=4$, $N=12$, $d=5$\,kB]{\includegraphics[width=0.49\columnwidth]{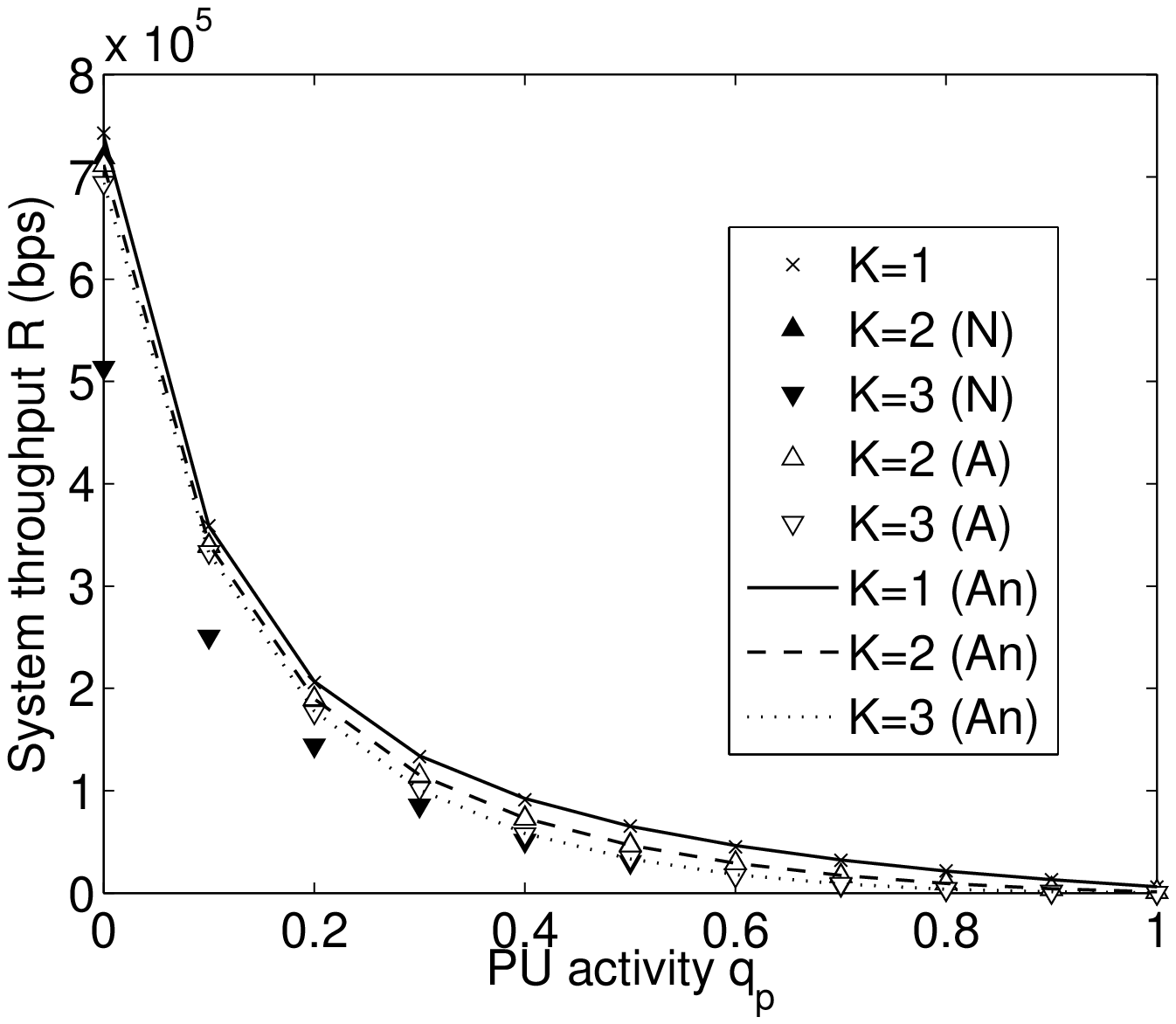}\label{fig:qp_a}}
\subfigure[$M=12$, $N=40$, $d=20$\,kB]{\includegraphics[width=0.49\columnwidth]{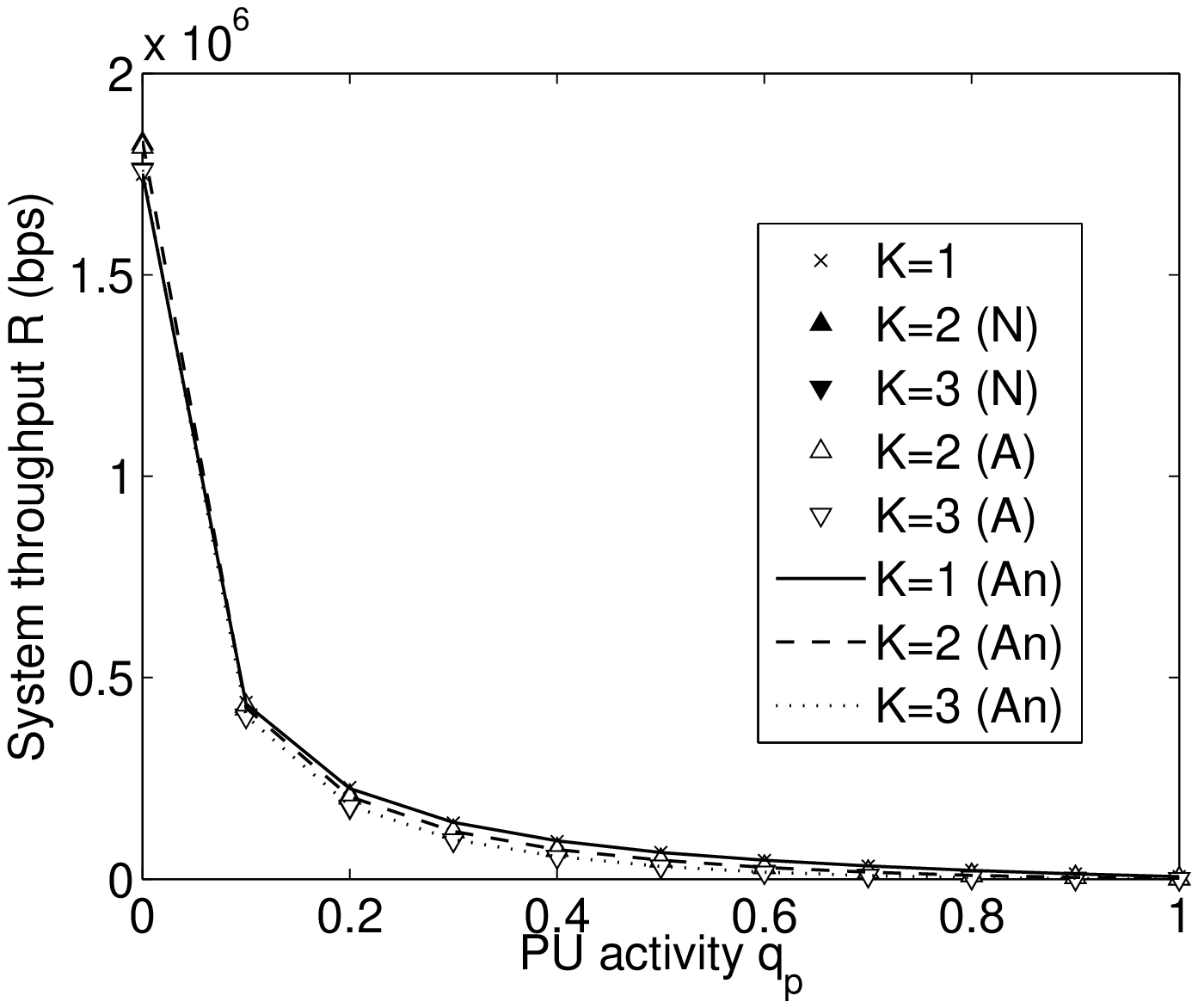}\label{fig:qp_b}}
\caption{Impact of PU activity level $q_p$ on system throughput for (a) small network and (b) large network; common parameters: $p=e^{-1}/N$, $C=200$\,kB, $T=1$\,ms, $T_s=0.1$\,ms, $\forall i$ $\beta(i)=1$, $\bar\gamma=0$\,dB, $\Omega=17.8$\,dB, $W=200$\,kHz, $p_d=0.9$, $p_f=0.02$; N:  $K$-only bonding (simulations), A: flexible bonding (simulations), An: analysis.}
\label{fig:impact_qp}
\end{figure}

We evaluate the impact of PU activity $q_p$ on the system throughput $R$. The results are presented in Fig.~\ref{fig:impact_qp}. The channel bonding scheme considered in the analysis is represented in Fig.~\ref{fig:qp_a} and Fig.~\ref{fig:qp_b} as `A'. For comparison we present the results for the $K$-only channel bonding scheme (represented in Fig.~\ref{fig:qp_a} and Fig.~\ref{fig:qp_b} as `N' and described in Section~\ref{sec:comparison_protocols}), where a new connection is dropped if and only if the number of available PU channels is less than $K$.

We observe that the impact of PU activity level $q_p$ on system throughput $R$ is strictly non-linear for every channel bonding order and network size. This behavior is due to the increased impact of PU preemptions on the active connections using virtual channels. Furthermore, increasing PU activity results in less throughput reduction. This is because for every channel bonding order just one active PU on any physical channel constituting a virtual channel causes disruption. Our result closely resemble simulation results presented in~\cite[Fig. 11]{bahl_sigcomm_2009}, where with an increase in the number of access points generating background traffic on TV UHF channels the throughput of WhiteFi platform decreases semi-exponentially. Both Fig.~\ref{fig:qp_a} and Fig.~\ref{fig:qp_b} present similar curve shapes. The largest impact is observed for low values of PU activity, i.e. $0\leq q_p \lessapprox 0.2$. Most importantly, we see that for these two network scenarios, small- and large-scale, the benefit of channel bonding diminishes as the PU activity increases.

For the small-scale network, the non-bonded system always outperforms the systems using channel bonding schemes. Furthermore, we observe that $K$-only bonding with $K=3$ yields the lowest throughput, irrespective of the PU activity level, and this difference is largest for low $q_p$ values. Note that both the flexible and $K$-only schemes exhibit the same performance for $K=2$ for the entire range of PU activity because each scheme bonds the same number of channels regardless of how many of them are available. This is true for any scheme in which $M$ is perfectly divisible by $K$. For other bond orders, flexible and $K$-only bonding schemes converge to the same value with an increase in $q_p$.

In contrast to the small-scale network scenario, we observe a benefit for channel bonding in comparison to a non-bonded system for $0\leq q_p\lessapprox 0.1$ in the large-scale network scenario. But as $q_p$ increases we observe that the benefit of bonding diminishes just as in the small-scale scenario and eventually the system experiences a slight degradation in performance as compared to the non-bonded system. This degradation is attributed to an increase in PU activity, which leaves more channels unutilized due to more preemption of bonded connections by PUs.

Finally, the analytical results obtained using the proposed mathematical model have been verified via simulations implemented in Matlab. We observe a perfect match between simulation and analytical results, as seen in Fig.~\ref{fig:qp_a} and Fig.~\ref{fig:qp_b}, where analysis is marked as `An'. This validates the correctness of the analytical model.

\subsection{Impact of User Pool Size on System Throughput}
\label{sec:resutls_ch_pool}

\begin{figure}
\centering
\subfigure[$M=4$, $T=2$\,ms, $q_p=0.1$]{\includegraphics[width=0.49\columnwidth]{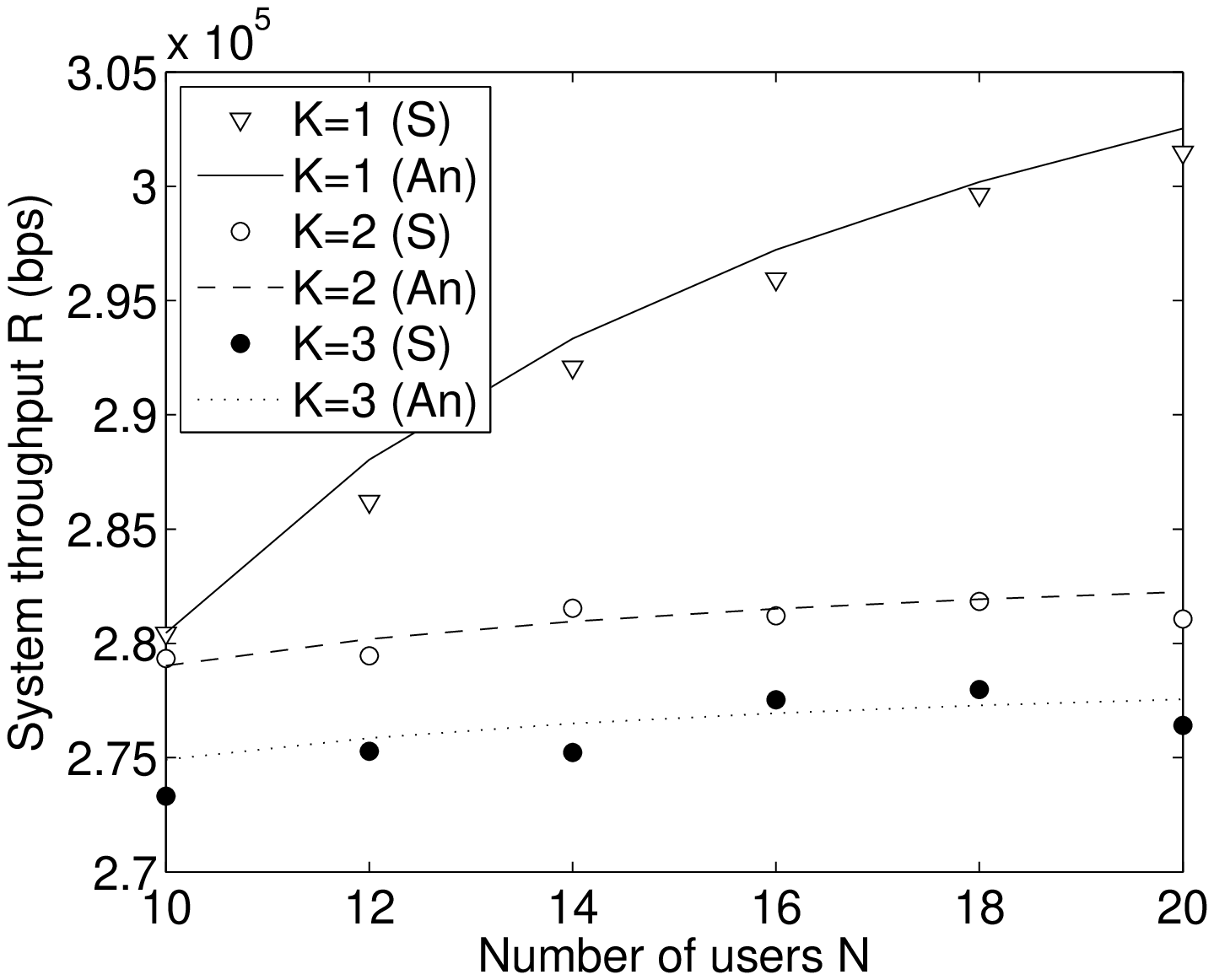}\label{fig:MN_a}}
\subfigure[$M=4$, $T=5$\,ms, $q_p=0.1$]{\includegraphics[width=0.49\columnwidth]{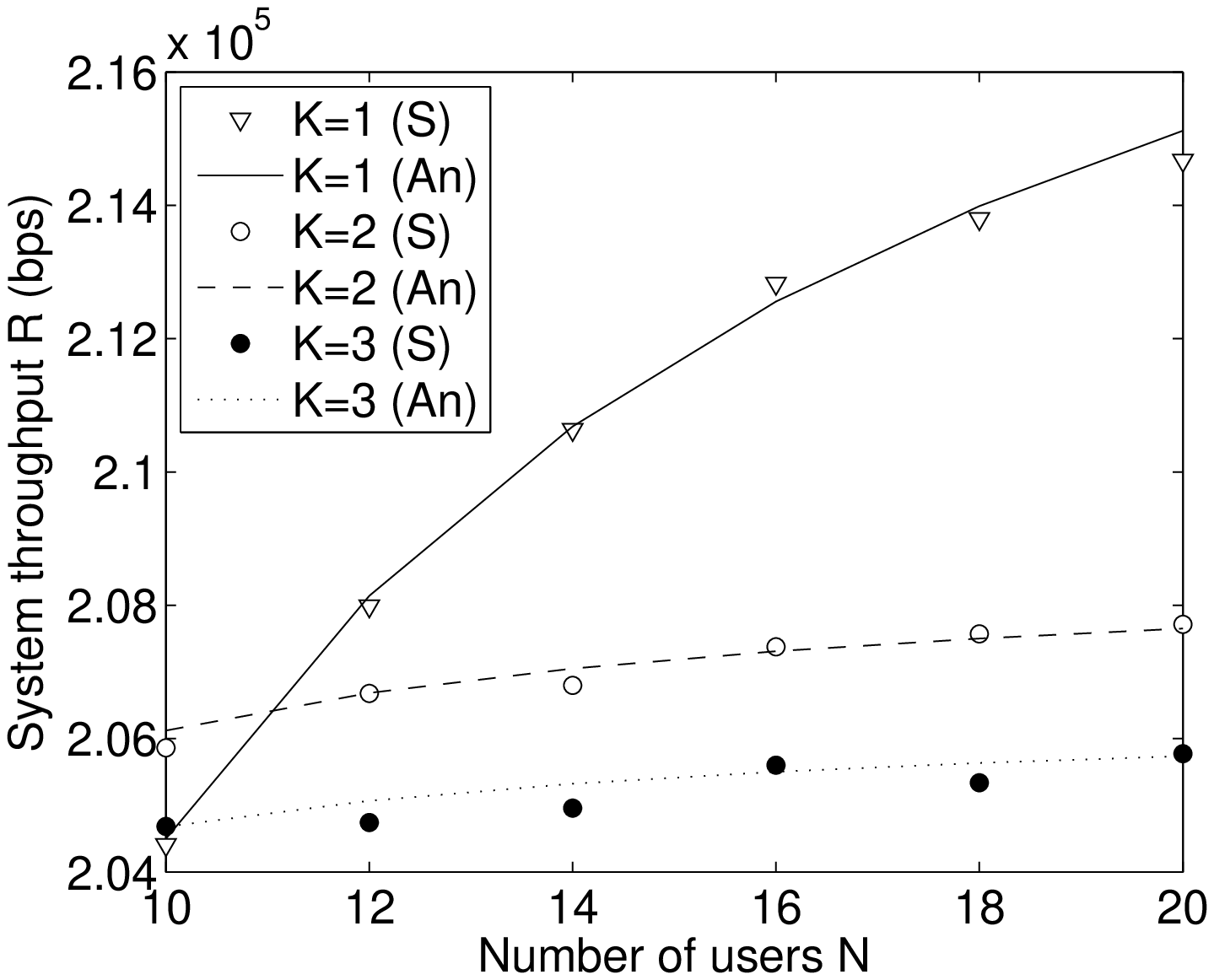}\label{fig:MN_b}}
\subfigure[$M=8$, $T=2$\,ms, $q_p=0.05$]{\includegraphics[width=0.49\columnwidth]{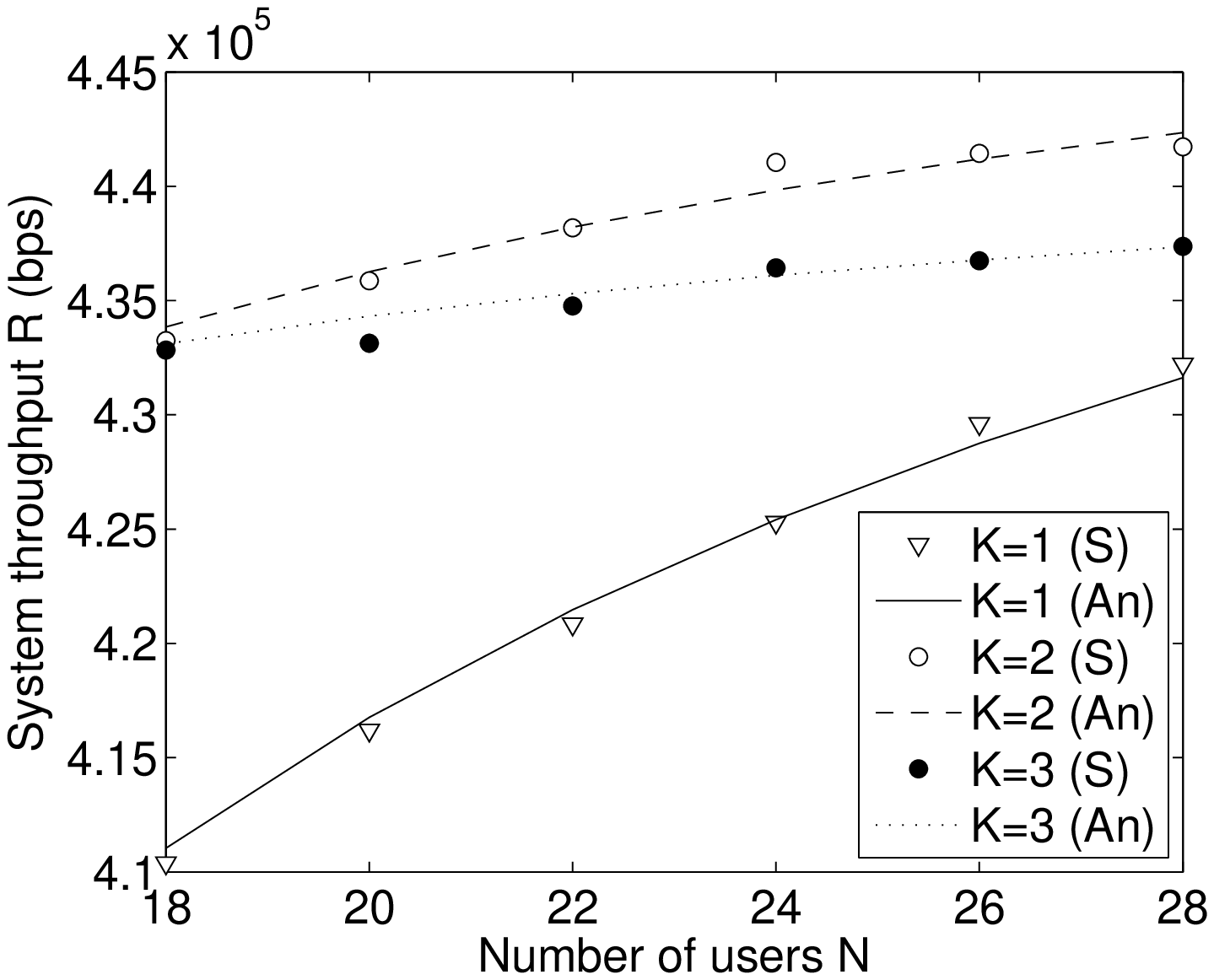}\label{fig:MN_c}}
\subfigure[$M=8$, $T=5$\,ms, $q_p=0.05$]{\includegraphics[width=0.49\columnwidth]{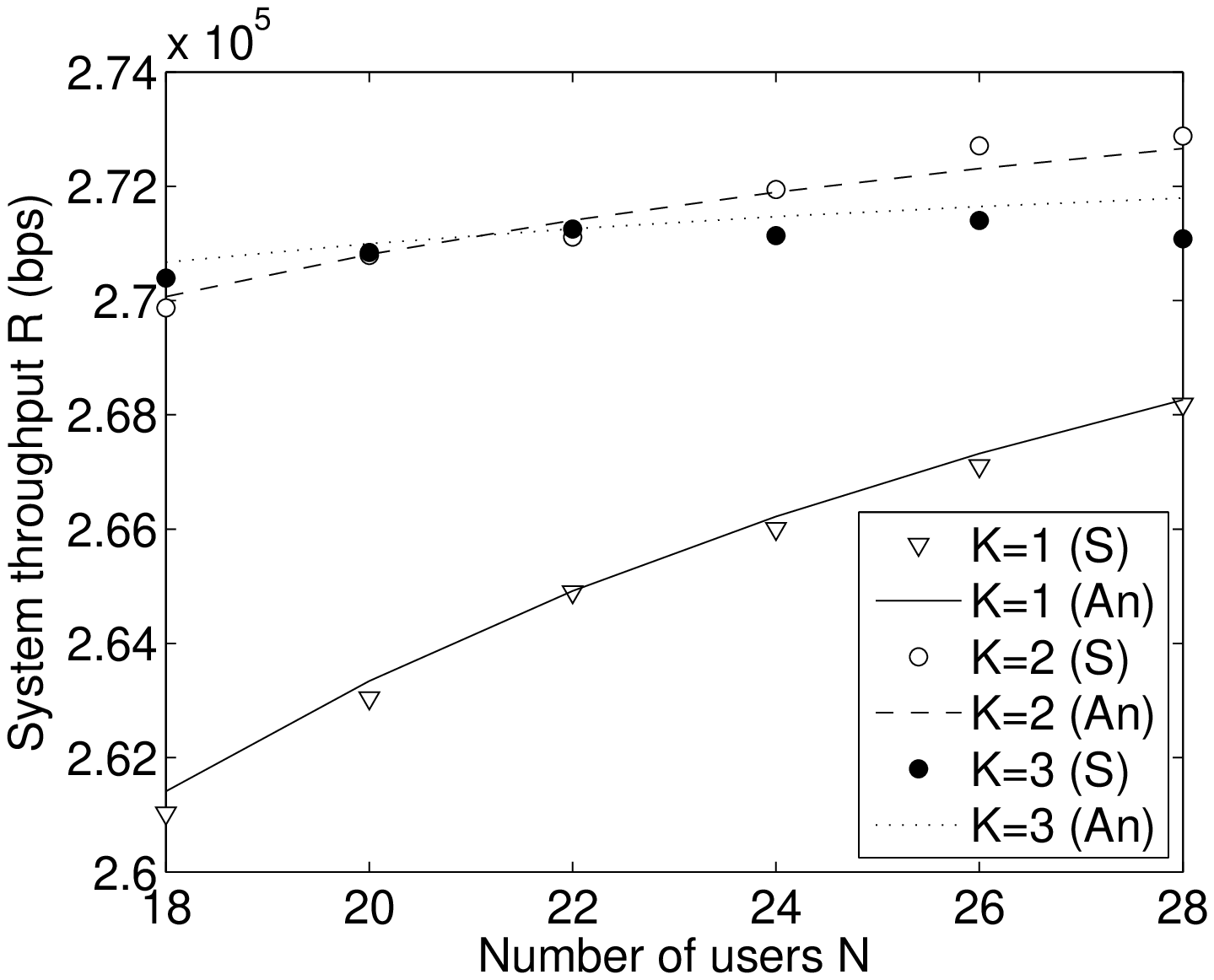}\label{fig:MN_d}}
\caption{Impact of user pool size $N$ on system throughput for (a) (b) small-scale network and large PU activity and (b) twice the size of the small-scale network and smaller PU activity, using flexible channel boding scheme; set of common parameters is the same as in Fig.~\ref{fig:impact_qp} assuming $d=1$\,kB; S: simulations, An: analysis.}
\label{fig:impact_MN}
\end{figure}

Exploring the performance of the protocol with a wider set of parameters, in contrary to Section~\ref{sec:result_pu_impact} we investigate the performance of the channel bonding protocol with two additional values for time slot length, i.e. $T=2$\,ms (i.e. time slot length in HSDPA standard) and $T=5$\,ms (i.e. time slot length in MobileWiMAX standard)~\cite[Sec. IV-A]{hassel_twc_2007}, two various PU activity levels, $q_p\in\{0.05,0.1\}$ (which are typical values for moderately used PU channels~\cite{wellens_phycom_2009}), and two network sizes, small with $M=4$ and large with $M=8$. Other parameters are the same as in Section~\ref{sec:result_pu_impact}. We divide the results into two groups: Fig.~\ref{fig:MN_a} and Fig.~\ref{fig:MN_b} present results for a small-scale network and large PU activity, and Fig.~\ref{fig:MN_c} and Fig.~\ref{fig:MN_d} present results for a larger-scale network and smaller PU activity.

Interestingly, as the number of users in the system increase, certain channel bonding orders yield better results than others for a fixed PU activity level. This phenomenon is scenario-dependent. In the case of the small-scale network, when the number of users per channel is relatively small, i.e. $N<12$, the system with $K\in\{2,3\}$ outperforms the system with no bonding, but only in the case of $T=5$\,ms time slot and small OSA user population size, compare Fig.~\ref{fig:MN_a} with Fig.~\ref{fig:MN_b}. As the number of users exceeds 12, the throughput from the system without channel bonding begins to exceed that obtained from higher bond orders for both small and large time slot lengths, compare again Fig.~\ref{fig:MN_a} with Fig.~\ref{fig:MN_b}. This is attributed to the increased number of collisions during the random access phase that higher bond order systems experience as compared to the non-bonded system. This occurs because more SU pairs remain unconnected. In other words, a smaller number of connected SU pairs, each occupying more resources with higher bond orders, is worse off than more connected SU pairs, occupying fewer resources, for the non-bonded system. This phenomenon becomes more exaggerated as the number of users in the system increase. In the case of the large-scale network, see Fig.~\ref{fig:MN_c} and Fig.~\ref{fig:MN_d}, the non-bonded system results in a drastically reduced throughput in comparison to the other bonded systems. With a small number of users in the system and $K=3$ bond order, throughput is maximized but again only for the case of $T=5$\,ms time slot, see Fig.~\ref{fig:MN_d}. As the number of users increases, the system with lower bond orders obtains higher throughput, as seen at $N=18$ when $K=2$ curve exceeds $K=3$ (Fig.~\ref{fig:MN_c} with a small time slot length) and at $N=22$ (Fig.~\ref{fig:MN_d} with larger time slot length). The intersection of the throughput curves for the non-bonded system and the scheme with $K=2$ occurs at a very large number of users because the rate in which throughput curves saturate is slower for networks with large channel pools.

Our analytical tool demonstrates the importance of determining channel bonding performance, since the intersection between different channel bond orders is strictly dependent on the scenario and is very difficult to deduce intuitively. Again, as in Section~\ref{sec:result_pu_impact}, the analytical results have been verified via simulations. A very good match between simulations and analysis further confirms the accuracy of the proposed mathematical model.

\subsection{Impact of Virtual Channel Throughput on System Throughput}
\label{sec:results_virtual_throughput}

\begin{figure}
\centering
\subfigure[$M=4$, $N=12$]{\includegraphics[width=0.49\columnwidth]{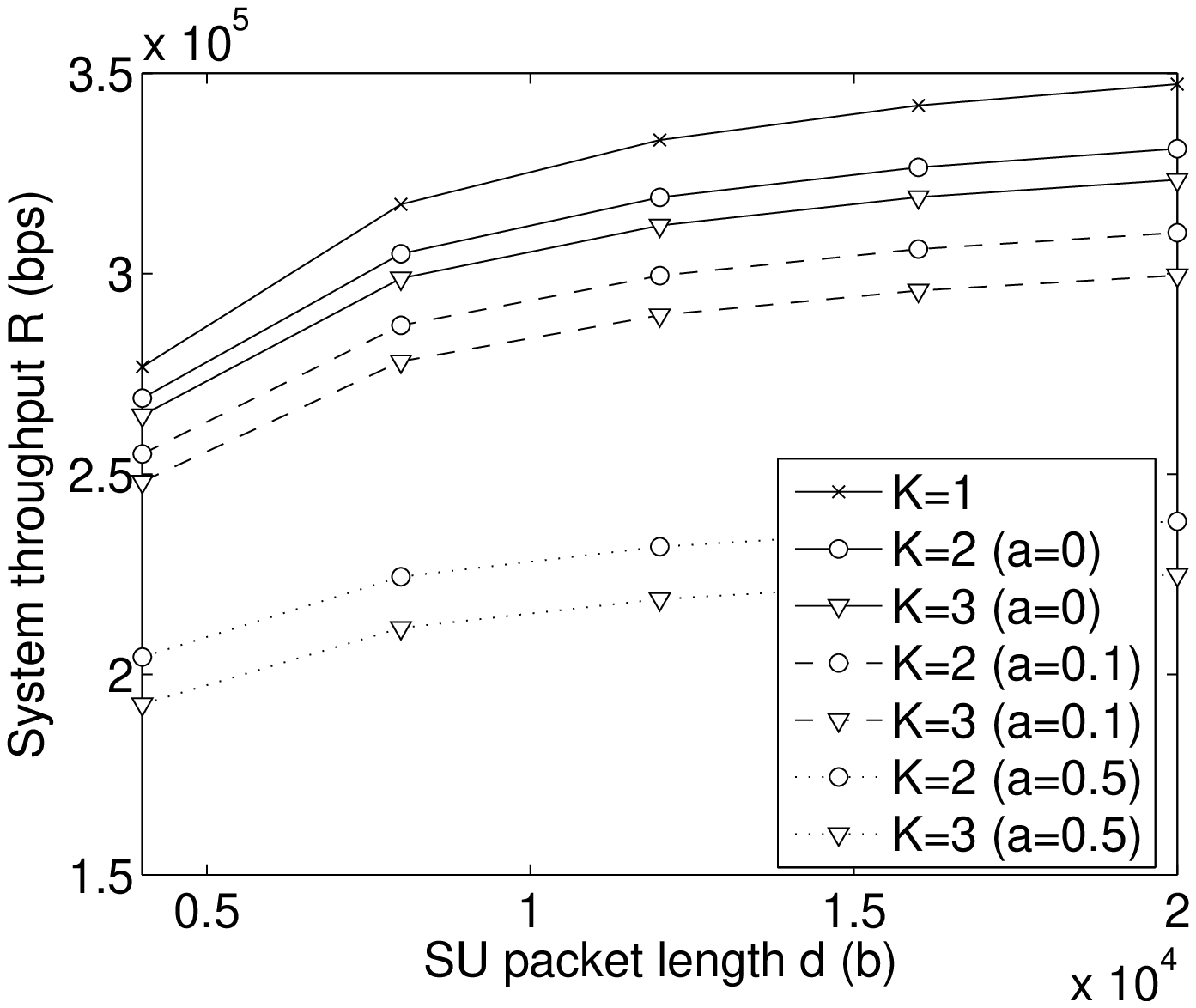}\label{fig:beta_a}}
\subfigure[$M=12$, $N=40$]{\includegraphics[width=0.49\columnwidth]{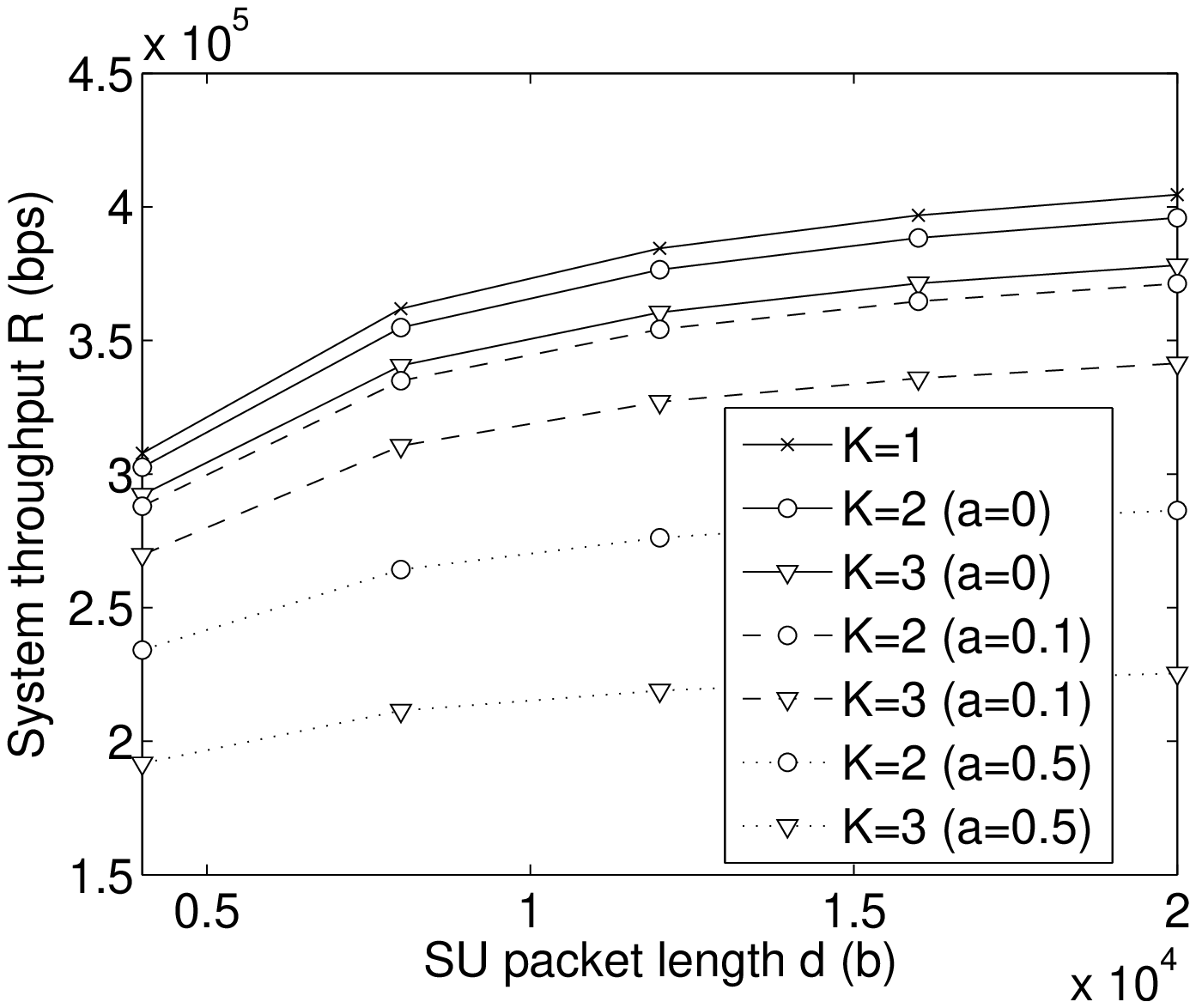}\label{fig:beta_b}}
\caption{Impact of virtual channel throughput on system throughput for (a) small network and (b) large network for two channel bonding schemes $K\in\{2,3\}$ and three virtual channel throughput reduction factors, using flexible channel boding scheme: (a) perfect throughput ($a=0$), (b) residual decrease ($a=0.1$) and (c) large decrease ($a=0.5$); set of common parameters is the same as in Fig.~\ref{fig:impact_qp} assuming $q_p=0.1$.}
\label{fig:impact_beta}
\end{figure}

We investigate the effect of virtual channel throughput on $R$ as shown in Fig.~\ref{fig:impact_beta}, where we investigate $R$ as a function of the SU frame size $d$. It has been suggested that the virtual channel throughput for higher bond orders is less than that of lower bond orders due to the increase in overhead when bonding~\cite{arslan_conext_2010}. We select three functions, as an example, to govern the way that virtual channel capacity is determined. These are: (a) perfect virtual channel throughput, for which $\forall k$ $\beta(k)=1$ (used in earlier two sections); (b) residual decrease in virtual channel throughput for which $\beta(k)=1/k^a$, $a=0.1$; and (c) large decrease in virtual channel throughput for which $\beta(k)=1/k^a$, $a=0.5$. The function $\beta(k)$ has been selected such that it decreases throughput exponentially and it penalizes channels with higher bond orders more than those with lower bond orders\footnote{Note, again that any function $\beta(k)$ can be selected for evaluation. The particular function in this paper was considered to represent the typical wireless transmission scenario~\cite{arslan_conext_2010}.}.

\begin{figure}
\centering
\subfigure[$M=4$, $N=12$]{\includegraphics[width=0.49\columnwidth]{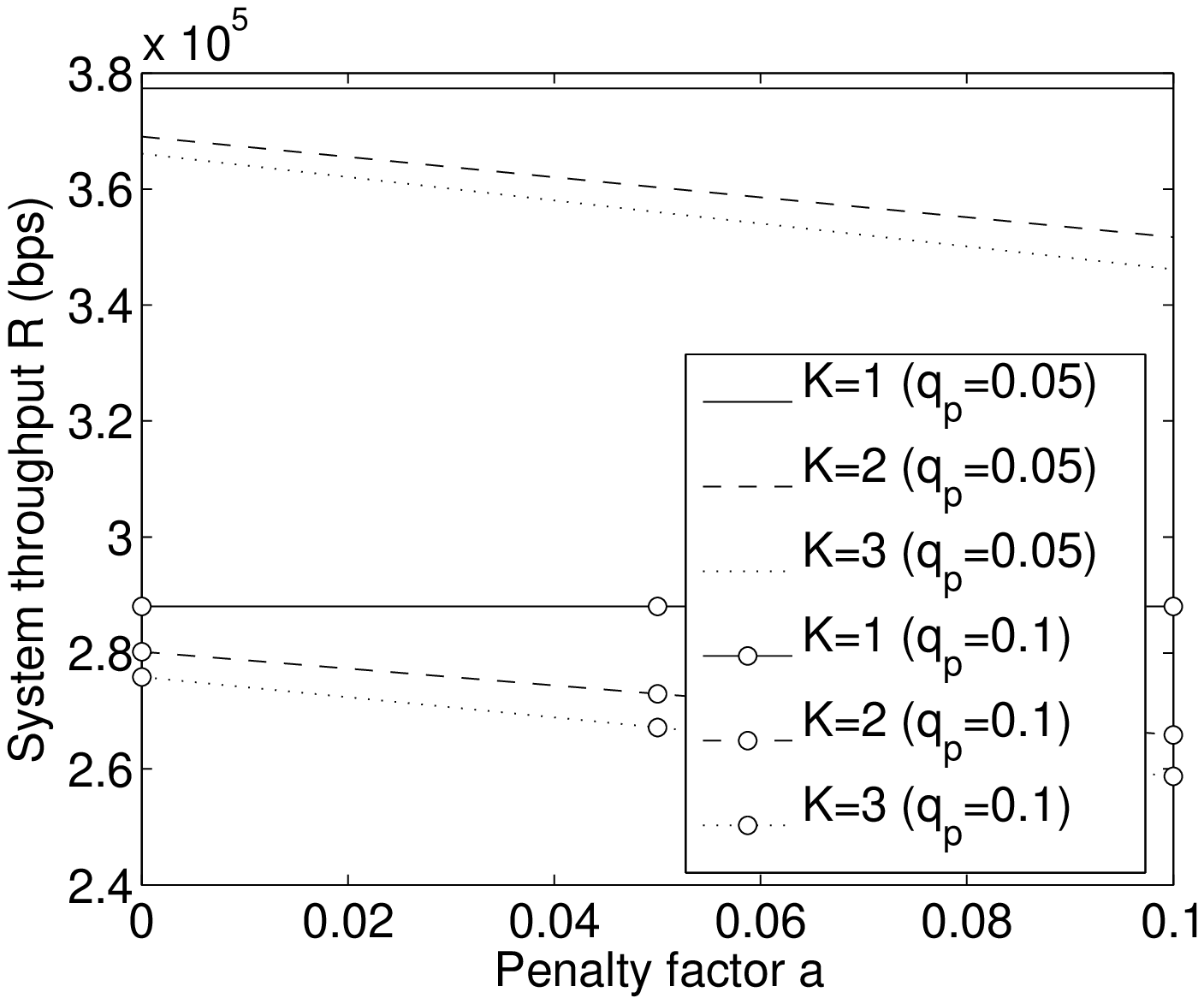}\label{fig:anicr_a}}
\subfigure[$M=8$, $N=24$]{\includegraphics[width=0.49\columnwidth]{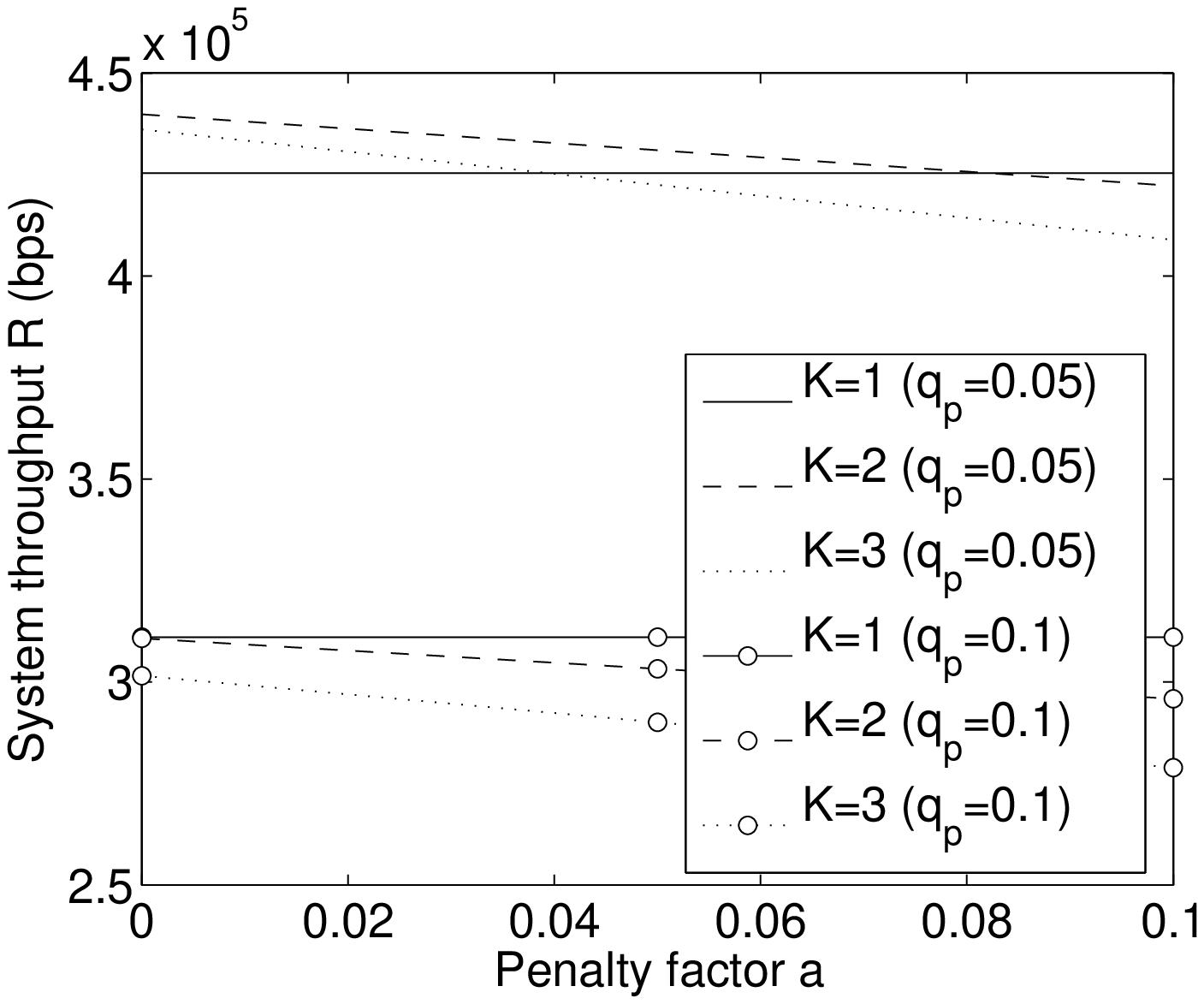}\label{fig:aincr_b}}
\caption{Impact of penalty factor on system throughput for (a) small network and (b) twice the size of the small network for two channel bonding schemes $K\in\{2,3\}$ and two values of PU activity $q_p\in\{0.05,0.1\}$ as a function of penalty factor, using flexible channel boding scheme; set of common parameters is the same as in Fig.~\ref{fig:impact_MN} with $d=1$\,kB and $T=2$\,ms as in Fig.~\ref{fig:MN_a} and Fig.~\ref{fig:MN_c}.}
\label{fig:impact_aincr}
\end{figure}

As the frame size becomes larger, the system throughput saturates. This observation is common for all considered scenarios, channel bond orders and virtual channel reduction factors. Note that the impact of frame length on system throughput in the context of non-opportunistic spectrum access multichannel MAC has been explained in much larger detail in~\cite[Sec. 4.4]{mo_tmc_2008}, with such discussion later extended to opportunistic spectrum access multichannel MAC in~\cite[Sec. 5.2.2]{park_tmc_2011} and~\cite[Sec. III-E1]{pawelczak_tvt_2009}. We refer to these papers for a longer explanation of this phenomenon. To summarize these findings, with longer frames SUs occupy PU channels for a longer time on average resulting in an improved system throughput. Furthermore, as frames/packets get longer the SUs do not have to contend for resources as often via the control channel, therefore the number of collisions on the control channel become smaller, which again translates into improved system throughput. The increase in system throughput with an increase in frame size occurs despite the increase in collision probability between PU and SU transmissions (note that the per time slot probability of collision between PU and SU transmission is independent of SU frame/packet size).

When frame sizes are relatively small, in the range of [10,20)\,kb, the impact of a residual penalty for virtual channels on system throughput does not drastically affect the performance as expected. However, with a larger penalty, system throughput performance greatly deteriorates, proving no additional benefit of channel bonding in an OSA network. The difference between system throughput with $a=0.1$ and $a=0.5$ is more profound for the small-scale than for the large-scale network scenario, compare Fig.~\ref{fig:beta_a} and Fig.~\ref{fig:beta_b}. Irrespective of the scenario, a bond order of $K=1$ (i.e. non-bonded system) outperforms the other bonding schemes for all SU frame lengths for every penalty factor. Please refer to the earlier explanation in Section~\ref{sec:result_pu_impact}.

In another set of experiments, we investigate the effect of increasing the severity of the penalty $a$ in $\beta(k)=1/k^a$ on virtual channels by observing the overall throughput for different channel bond orders and two different PU activity levels, $q_p\in\{0.05,0.1\}$ with results presented in Fig.~\ref{fig:impact_aincr}. We observe an approximately linear decrease in system throughput with an increase in the penalty factor $a$. For a small-scale network, see Fig.~\ref{fig:anicr_a}, as well as another network which is twice the size of the small-scale network, see Fig.~\ref{fig:aincr_b}, even a very small penalty for the virtual channel results in system throughput loss as compared to the non-bonded system, for example compare $K=1$ and $K=2$ for $a\lessapprox 0.02$ and both values of $q_p$. In the small-scale scenario, the throughput of higher bond orders is worse than that of the non-bonded system, as shown in Fig.~\ref{fig:qp_a}. However, in the case of the larger network, see Fig.~\ref{fig:aincr_b}, the non-linearities of virtual channel throughput are better accommodated when PU activity is low (again, in our case $q_p=0.05$).  Hence the bonded systems provide for an improvement in throughput in the range of $0\leq a \lessapprox 0.04$. This proves that there is a benefit of channel bonding even when the OSA network is unable to exploit the full theoretical capacity provided by the virtual channel. Finally, we add that (not shown in the Fig.~\ref{fig:impact_aincr} due to space constraints) as the PU activity increases, the effect of the penalty function becomes marginal as preemption dominates incoming SU connections.

\subsection{Impact of Virtual Channel Disruption Resolution Strategy on System Throughput}
\label{sec:results_pu_distruption}

Thus far we assume that once the PU is detected on any of the virtual channels the transmitted frame is lost. This approach serves as a lower bound on the channel bonding MAC protocol performance, refer to Section~\ref{sec:primary_user_detection} for details. In this section we will relax this assumption and consider an example of a more flexible frame disruption resolution strategy.

Specifically, we extend our simulation platform and consider a strategy denoted as Channel Switching (S), where the OSA network frame on the arrival of the PU (on any of the physical channels) is being switched to other free PU channels\footnote{We emphasize that we consciously do not consider frame disruption resolution strategies which exclude physical channels occupied by PUs from the virtual channel, as this would result in an unfair comparison with the flexible case of the OSA channel bonding MAC protocol. This is because the frame transmission rate in the flexible case does not change dynamically, i.e. $q(k)$ being a function of bonding size $K$ is kept constant throughout the whole network operation.}. In other words, any frame transmission that uses $K$-bonded channels must be switched to another set of $K$ unoccupied physical channels. Thus, the frame is lost only when not enough physical channels are found for switching. As S strategy is equivalent to B$_0$S$_1$ in~\cite{park_tmc_2011} we refer to~\cite[Theorem 1]{park_tmc_2011} proving that distributed channel switching can be performed by the OSA network without additional signaling transmission. Furthermore, \cite[Sec. 4.3.2]{park_tmc_2011} describes in detail how the OSA multichannel MAC protocol with DCC resolves collisions when multiple sender/receiver pairs decide to switch to the same pool of physical channels. For notational convenience, in the remainder of this section the fundamental frame disruption resolution strategy (considered in the analysis) is denoted as A.
\begin{figure}
\centering
\subfigure[$q_p=0.1$, $d=5$\,kB]{\includegraphics[width=0.49\columnwidth]{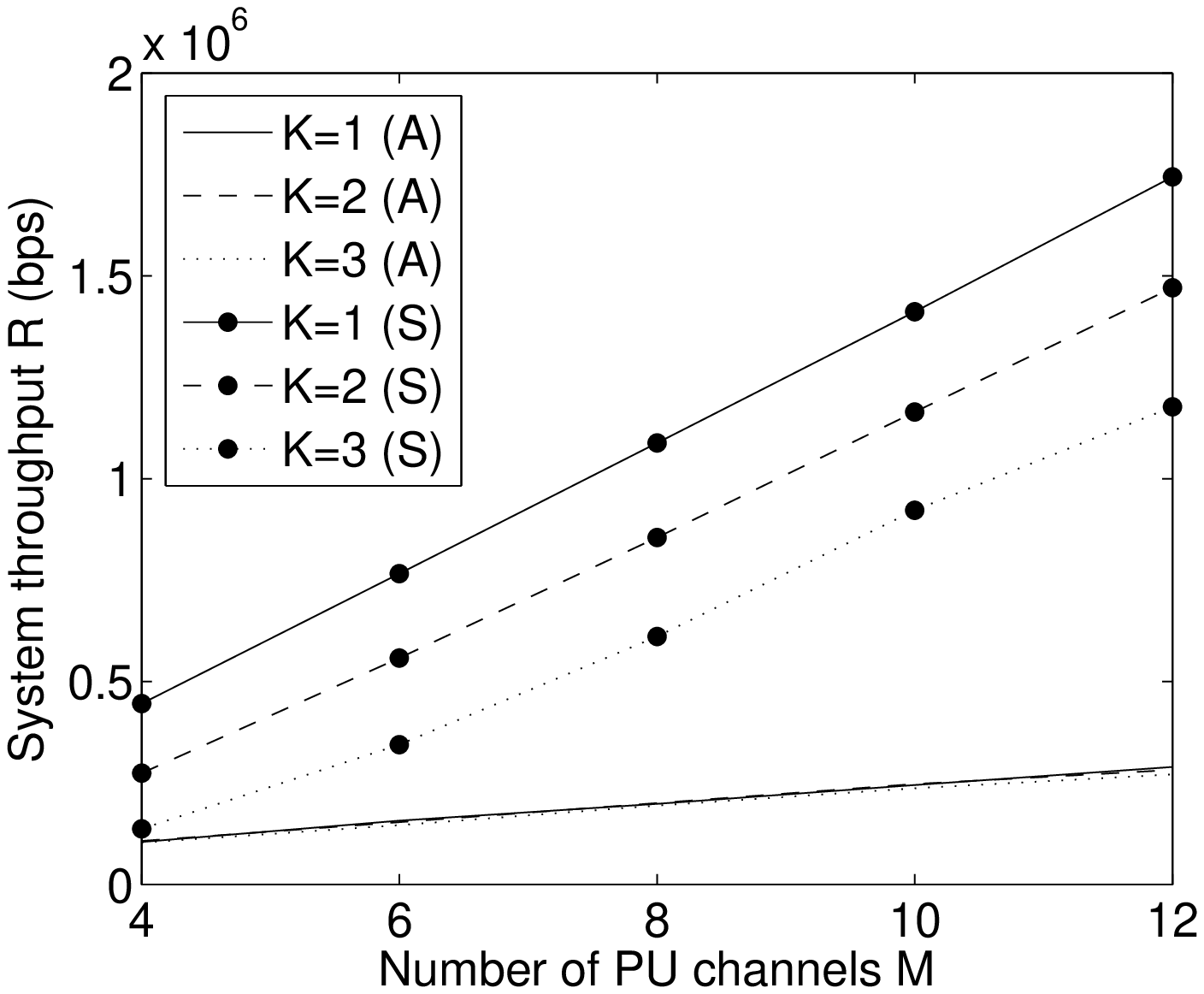}\label{fig:M_thr_qp01}}
\subfigure[$q_p=0.1$, $d=5$\,kB]{\includegraphics[width=0.49\columnwidth]{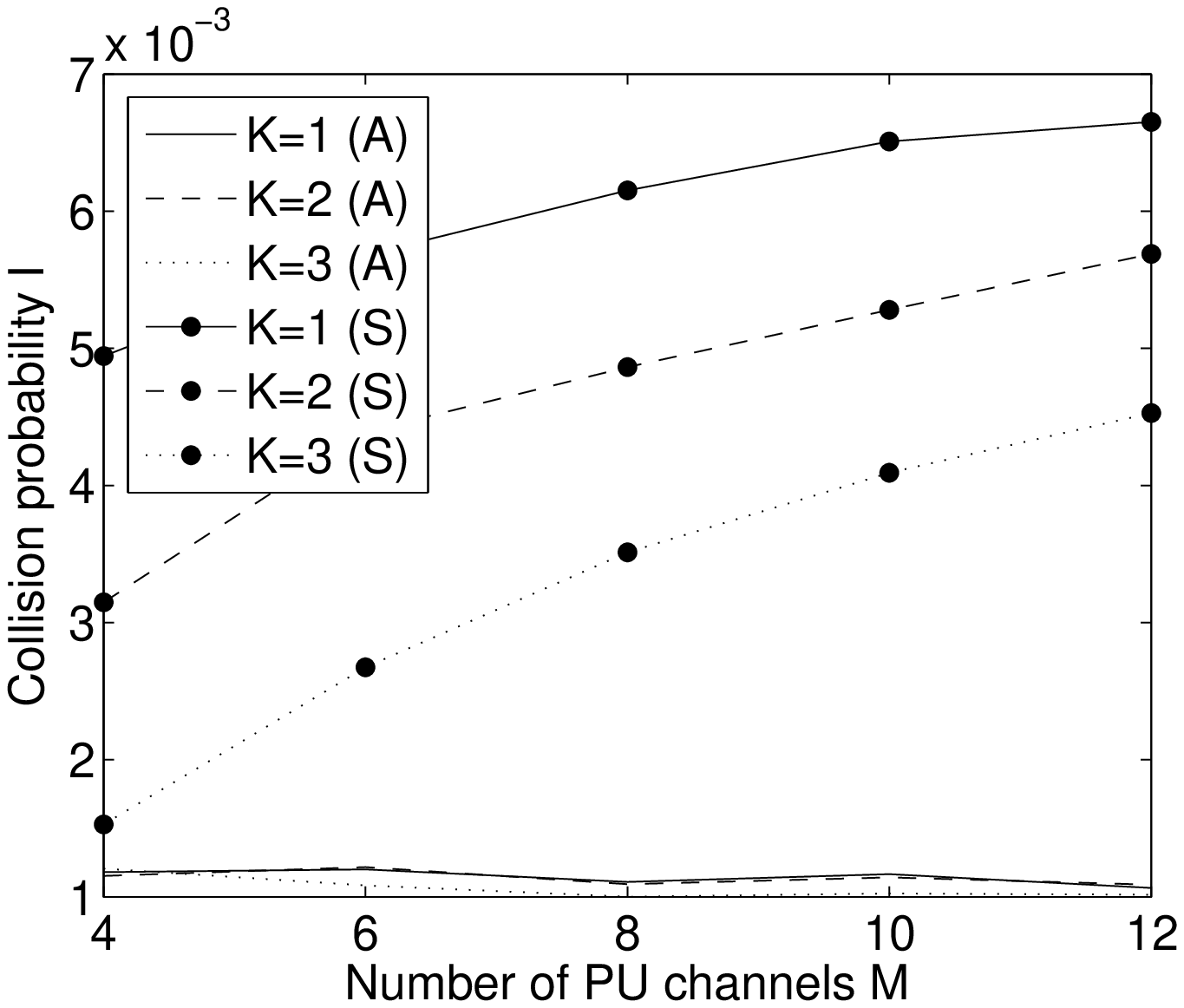}\label{fig:M_col_qp01}}\\
\subfigure[$q_p=0.3$, $d=20$\,kB]{\includegraphics[width=0.49\columnwidth]{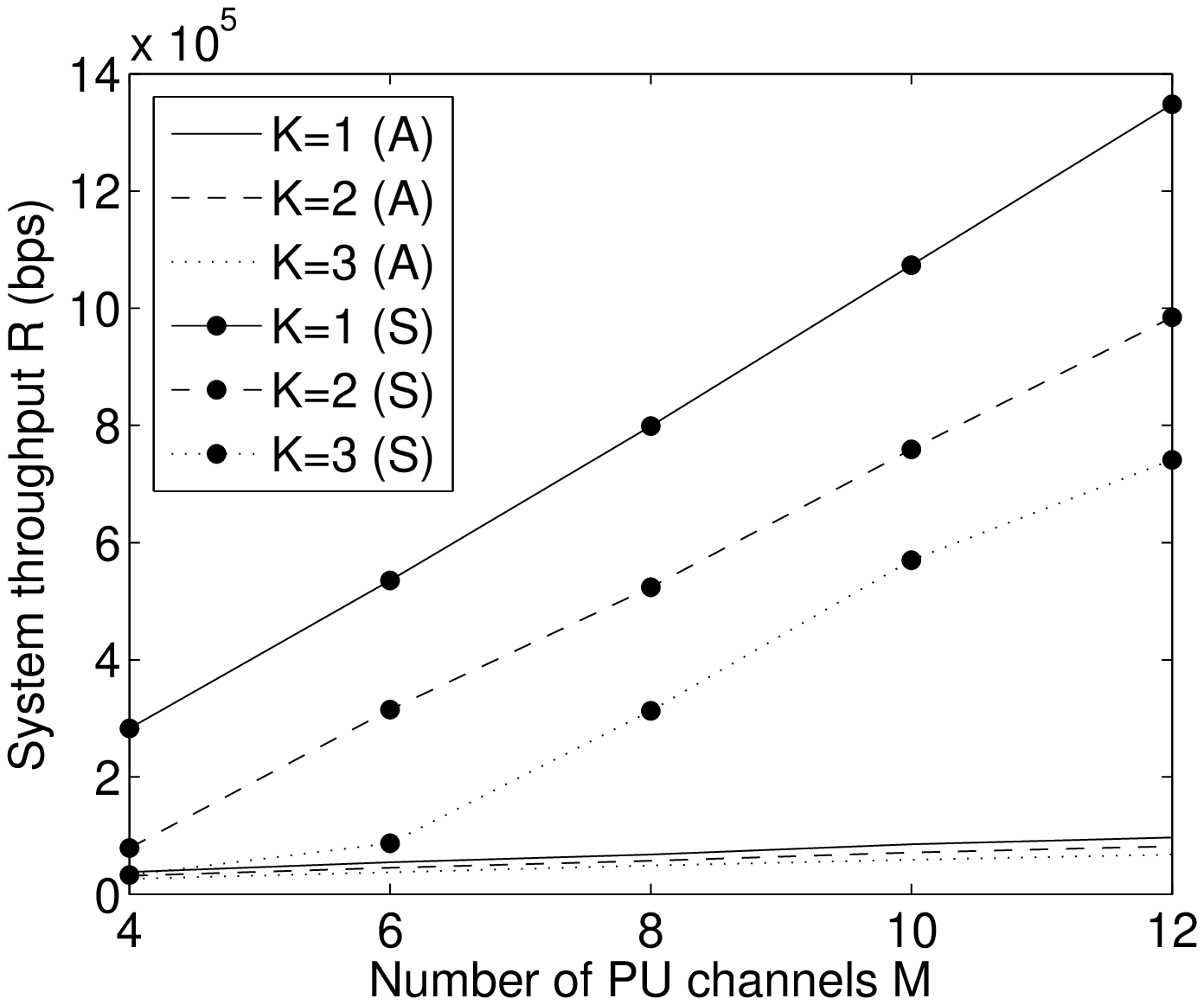}\label{fig:M_thr_qp03}}
\subfigure[$q_p=0.3$, $d=20$\,kB]{\includegraphics[width=0.49\columnwidth]{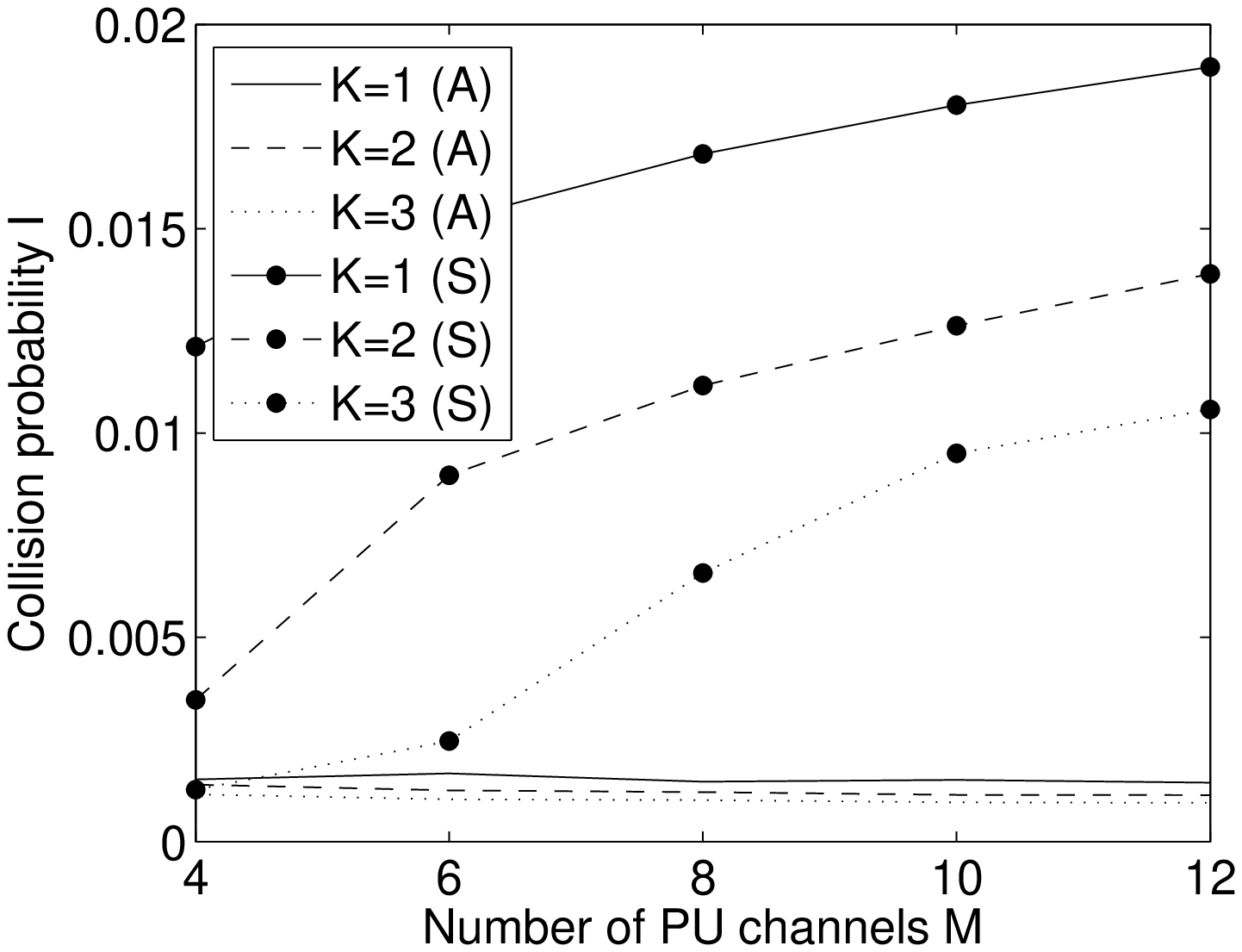}\label{fig:M_col_qp03}}
\caption{Impact of PU channel pool size on system throughput and collision probability for (a), (b) $q_p=0.1$, $d=5$\,kB and (c), (d) $q_p=0.3$, $d=20$\,kB for two OSA frame disruption resolution strategies, fundamental (lower bound) (A) and switching (S), three different bonding levels $K\in\{1,2,3\}$, $N=2M$ and channel switching time $T_p=100$\,$\mu$s; set of common parameters is the same as in Fig.~\ref{fig:impact_qp}.}
\label{fig:impact_M_col}
\end{figure}

In this section, in addition to throughput we investigate collision probability, $I$, defined as the probability of using a physical channel when the PU is present. Note that due to the channel switching process, to keep the comparison with flexible OSA channel bonding MAC fair, we virtually prolong the frame length by the channel switching delay $T_p$ such that $q(k)=\{C(T-T_s+T_p)/d\}k\beta(k)$. The results are presented in Fig.~\ref{fig:impact_M_col} where for fixed $N=2M$, $T_p=100$\,$\mu$s, and two values of $q_p$ and $d$ we observe the performance metrics of interest.

It is immediately observed that with the S strategy the OSA network with the channel switching performs much better than with the A strategy, see Fig.~\ref{fig:M_thr_qp01} and Fig.~\ref{fig:M_thr_qp03} (the increase is in orders of magnitude in comparison to strategy A). But on the other hand we also observe a large increase in the respective collision probabilities with the introduction of strategy S, see Fig.~\ref{fig:M_col_qp01} and Fig.~\ref{fig:M_col_qp03}. This result clearly shows that there is a trade-off between frame disruption resolvability and collisions induced to PUs by OSA operation. The larger the OSA frame and and the larger the PU activity per channel, the larger the collision rate, see Fig.~\ref{fig:M_col_qp03} where almost 2\% collision probability is reached (which is again in the orders of magnitude larger than in the A strategy). Interestingly, with increase in $K$ the collision probability rate decreases. This is because with a large channel bonding pool, there is less probability that free channels can be found for switching and the frame is dropped, in-turn reducing probability of further collisions with the PU.

We want to emphasize that with the S strategy, the conclusions obtained for OSA channel bonding MAC with the A strategy still hold. Mainly, the order of bonding preference is preserved and in most cases $K=1$ outperforms other bonding schemes in terms of throughput.

\subsection{Impact of Varying Channel Access Control and Prioritization on System Throughput and Fairness}
\label{sec:results_fairness}

\subsubsection{Non-Uniform PU Channel Utilization}
\label{sec:non-uniform_fairness}

In previous sections we have assumed that the channels are used equiprobably by each PU. In this case, the OSA network channel selection strategy does not have an effect on the MAC protocol performance as each selected channel has the same probability of being disrupted by the PU. However, as the channels become non-uniformly used by PUs, the selection of a successive channel for each new bond might have a profound effect on the OSA network performance. Therefore in this section we investigate the effect of channel selection strategies on the performance of the channel bonding MAC protocol with a non-uniform distribution of PU activity over the PU channel set.

Intelligent channel bonding strategies, assuming side-note information on PU channel properties were investigated in~\cite[Sec. 5.2]{kone_imc_2010} (denoted therein as frequency bundling), based on a large data set of measured PU channels. We will also consider such a system in the context of our work. Specifically, we will assume that the OSA network is aware of non-uniform PU channel usage and knows the first moment of PU channel distribution for every channel $i\in\{1,\cdots,M\}$. As an example, for mathematical tractability and without loss of generality, we assume that channel $i$ is used by PU with probability
\begin{equation}
q_p^{(i)}=\frac{q_pM}{\sum_{j=1}^{M}j^{-A_q}i^{A_q}},
\label{eq:chan_imb}
\end{equation}
where $A_q$ denotes the per channel PU activity imbalance factor. Note that $A_q=0$ denotes equal channel utilization by the PU for each channel and $A_q>0$ denotes an exponential increase in channel utilization with increasing $i$. Equation~(\ref{eq:chan_imb}) allows for a fair comparison of different channel selection strategies against uniform PU channel usage, as the mean of set $\{q_p^{(i)},\cdots,q_p^{(M)}\}$ for any $A_q$ is equal to the mean channel utilization in the case of uniform PU activity over all PU channels, i.e. $q_p$. Note, however, that in (\ref{eq:chan_imb}) $q_p$ must be selected such that it does not violate PU channel usage properties, i.e. $\forall i$ $0\leq q_p^{(i)}\leq1$.

In the simulation experiment we consider two channel selection strategies: random (R) channel selection and Least-used (L) channel selection. In case of the L channel selection strategy the OSA network, knowing how channels are used by the PU, always selects channels that are on average lowest used by the PU for each new bond (to minimize probability of collision with the PU and prolong transmission time). Apart from observing the OSA network throughput, for a considered distribution of channel usage by the PU, we also observe how uniformly channels were used by the SUs in the OSA network. For that we adopt a common fairness indicator denoted as Jain's index, defined as~\cite[Sec. 3]{jain_tr_1984}
\begin{equation}
F=\frac{\left(\sum_{i=1}^{M}f_i\right)^2}{M\sum_{i=1}^{M}f_i^2},
\label{eq:fairness}
\end{equation}
where $f_i$ is the usage of channel $i$ by the OSA network.

\begin{figure}
\centering
\subfigure[$M=4$, $N=12$, $d=5$\,kB]{\includegraphics[width=0.49\columnwidth]{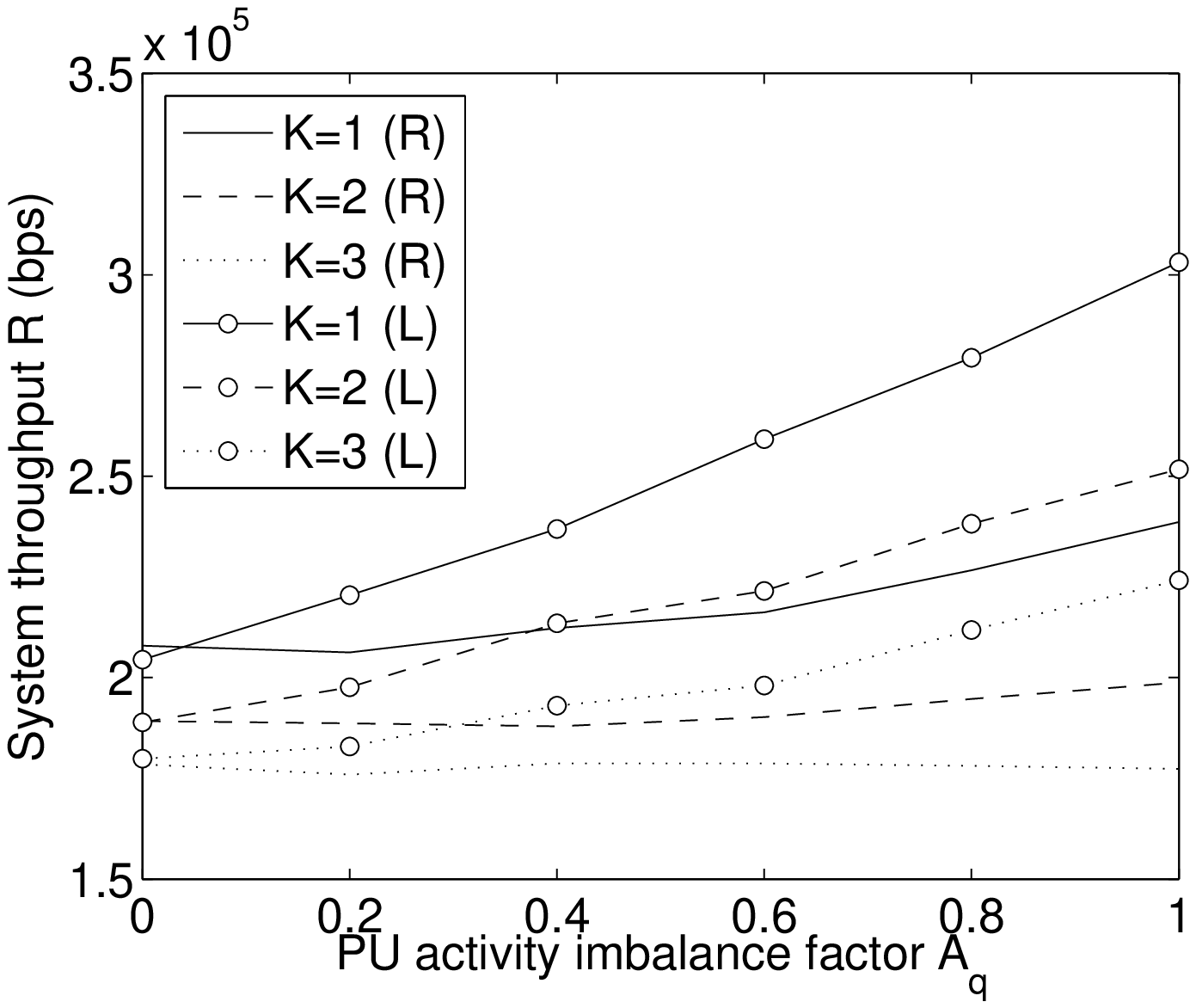}\label{fig:Aq_thr_small}}
\subfigure[$M=4$, $N=12$, $d=5$\,kB]{\includegraphics[width=0.49\columnwidth]{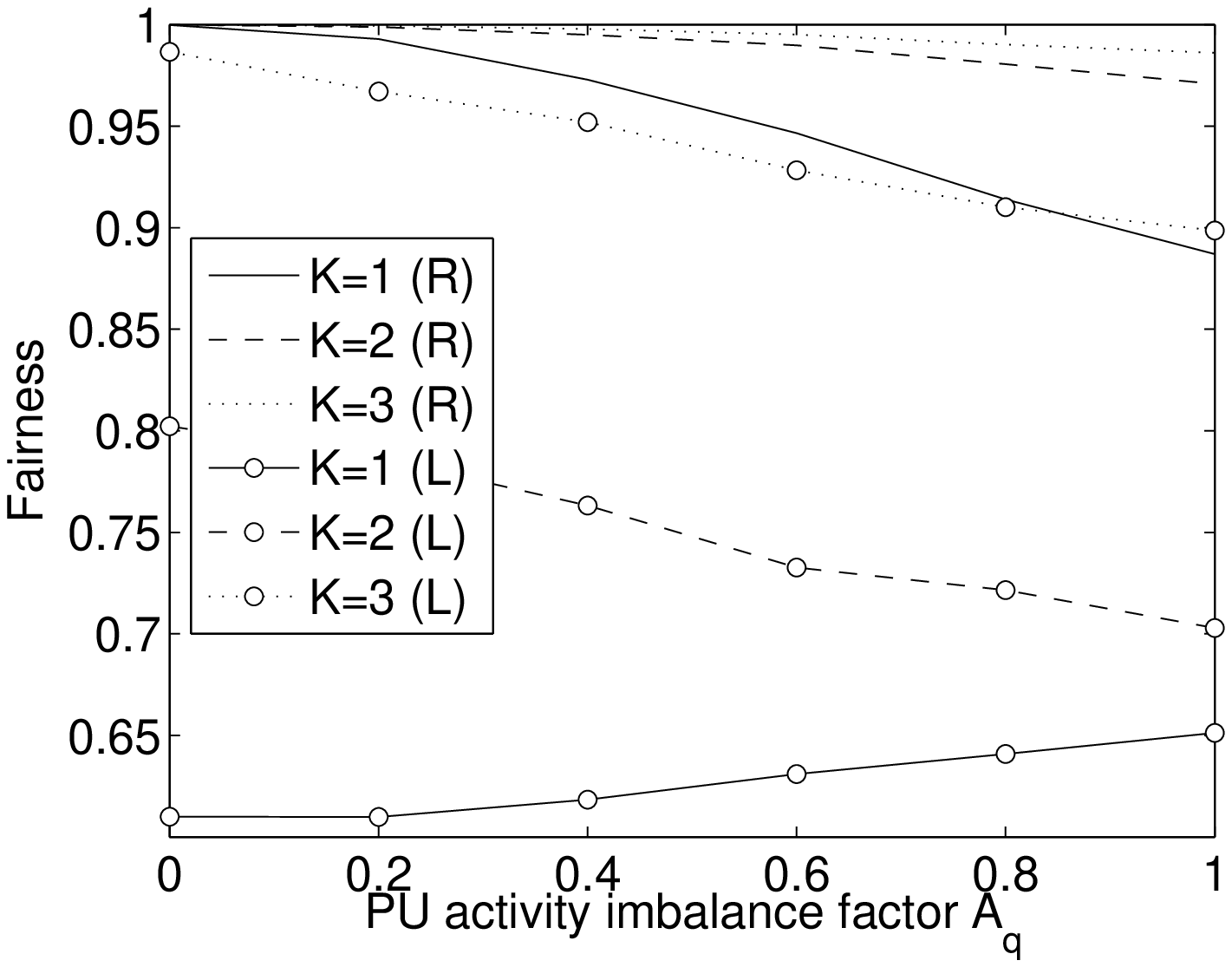}\label{fig:Aq_fai_small}}\\
\subfigure[$M=12$, $N=40$, $d=20$\,kB]{\includegraphics[width=0.49\columnwidth]{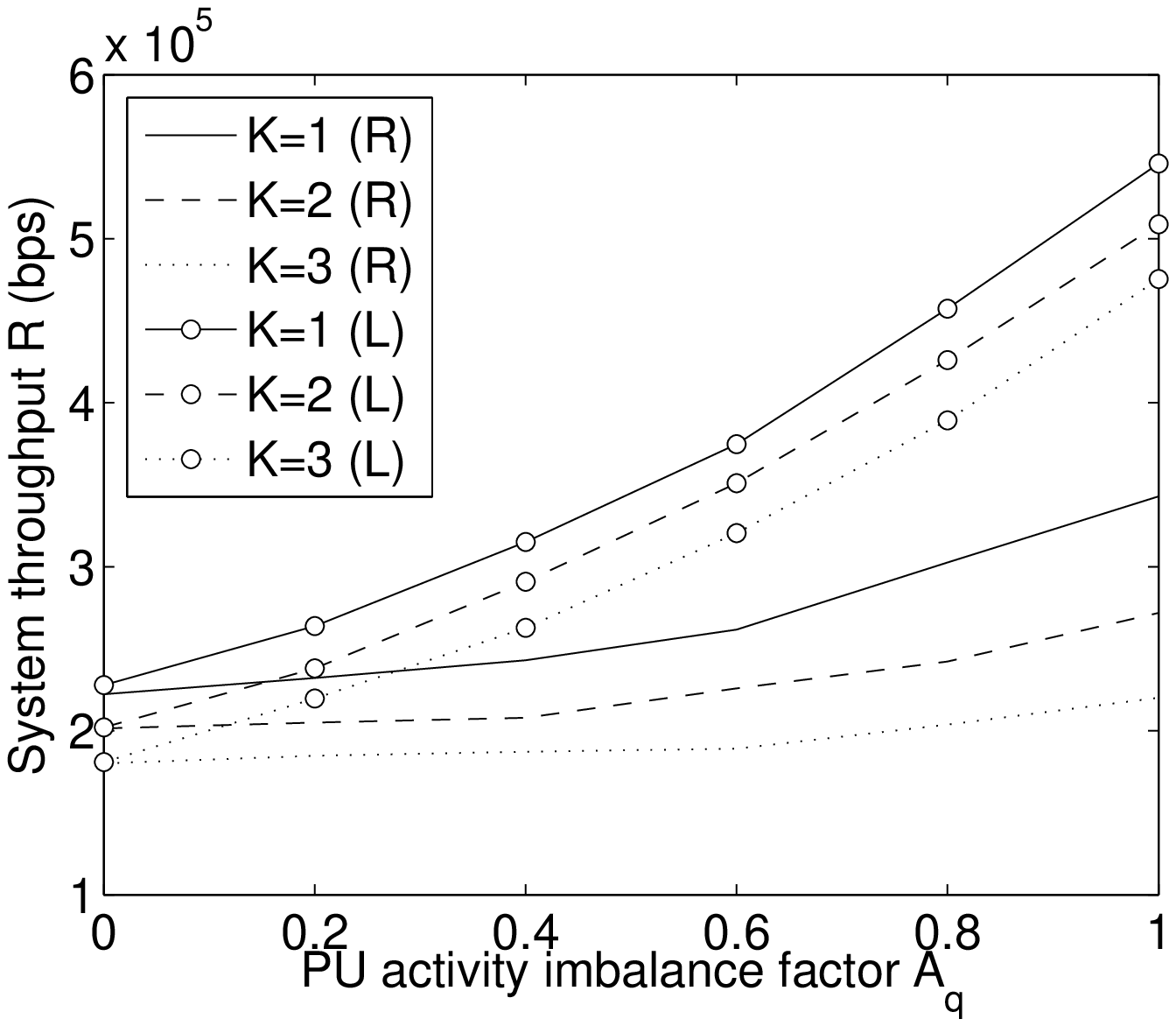}\label{fig:Aq_thr_large}}
\subfigure[$M=12$, $N=40$, $d=20$\,kB]{\includegraphics[width=0.49\columnwidth]{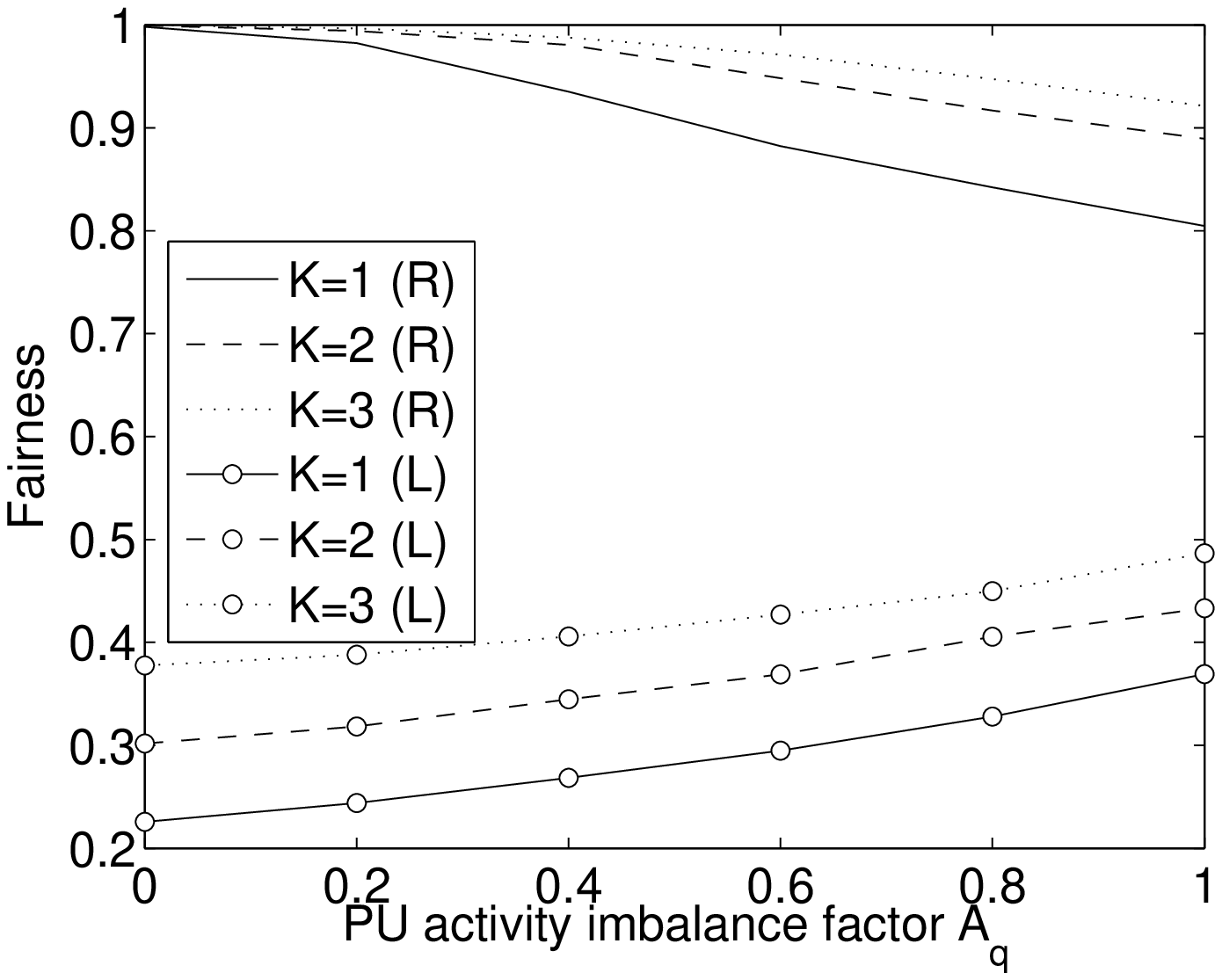}\label{fig:Aq_fai_large}}
\caption{Impact of PU channel usage imbalance on system throughput and fairness for (a), (b) small network and (c), (d) large network for two channel selection strategies, random (R) and least used (L), three different bonding levels $K\in\{1,2,3\}$ and $q_p=0.2$; set of common parameters is the same as in Fig.~\ref{fig:impact_qp}.}
\label{fig:impact_Aq}
\end{figure}

We present the results in Fig.~\ref{fig:impact_Aq}. The L channel selection strategy significantly outperforms the R channel selection in terms of obtained throughput, see Fig.~\ref{fig:Aq_thr_small} and Fig.~\ref{fig:Aq_thr_large}. The gain from using the L channel selection increases as $A_q$ becomes larger. This phenomenon holds irrespective of the channel bonding size $K$ and is more evident as network size increases. An obvious result is that for the L channel selection strategy the fairness is much lower than that of the R strategy since the R strategy has almost a perfect fairness of 1, especially for the small network scenario (see Fig.~\ref{fig:Aq_fai_small} and compare with Fig.~\ref{fig:Aq_fai_large}). However, a more interesting (non-obvious) result is that with an increase of bonding size $K$ the fairness of the L strategy increases. This is due to more channels being used by the MAC protocol simulatneously with a larger bonding order. Furthermore, for $K>1$ fairness decreases as $A_q$ increases. As the PU activity distribution becomes skewer, i.e. $A_q$ becomes large, the fairness of both the L and R strategies converge. This is especially visible in Fig.~\ref{fig:Aq_fai_large} in the case of a large OSA network because most of the PU activity is concentrated over a small pool of channels, while in other cases the channels are sporadically used by PUs. Interestingly, varying the channel selection strategy does not change the conclusions obtained earlier, e.g. $K=1$ reaches highest throughput followed by $K=2$ and then $K=3$.

\subsubsection{SU Channel Access Prioritization}
\label{sec:non-uniform_fairness}

In addition, we consider a fundamentally different channel bonding OSA MAC protocol, where priorities within the SU network are allowed. In particular, we consider a flexible channel bonding MAC, where each SU is able to send two types of frames: (i) regular and (ii) high priority.

The priority-enabling flexible channel bonding MAC works as follows. Each newly generated SU frame is a high priority frame occurring with probability $p_h$. An active, regular SU connection is halted and displaced on the arrival of a high priority SU connection when all channels are already occupied by active SU connections. A set of $B\in\{0,\cdots,M\}$ halted connections in total is allowed, where $B$ refers to the size of the buffer. Note that while in the real-life the most realistic setup for the OSA network is $B=M$ (as buffering occurs by halting the existing connection of one SU transmitter/receiver pair by means of observing connection requests on the control channel~\cite[Sec. 4.3.3]{park_tmc_2011}) in numerical evaluation we will also consider the effect of $B<M$ on the system throughput. Furthermore, for consistency with earlier results presented in this paper we do not consider buffering of SU connections on the event of preemption by PU connection, noting that the detailed consideration of multichannel (non-bonding) OSA MAC protocols with frame buffering, considering PU preemptions, has already been considered in~\cite[Sec. 4.3.3]{park_tmc_2011}. If more than one channel is freed from PU occupancy or SU transmission (of any kind), then one of the buffered connections resumes transmission until either the frame is successfully transmitted or preempted by the presence of a PU. The connections to be resumed are selected randomly from the pool in the buffer and per time slot one connection reconnect is possible (for consistency with the connection arrangement policy on the control channel). If during a single time slot new connection wants to access the channel pool and there is an existing connection to be resumed, the new connection gets priority over the buffered connection only if it is a high-priority connection. For regular connections, the buffered connection gets priority to connect to the channel. Furthermore, note that existing connection in the buffer does not contend again for channel access through the control channel. When one of the SU connections needs to be buffered, it is also selected randomly from the set of existing connections. In practical OSA this can be done by, e.g., implementing a rule of buffering first the connections of a SU sender that have been most buffered so far (each SU can easily track which connections were buffered by observing the control channel, refer to~\cite[Theorem 1]{park_tmc_2011} by analogy).

\begin{figure}
\centering
\subfigure[]{\includegraphics[width=0.49\columnwidth]{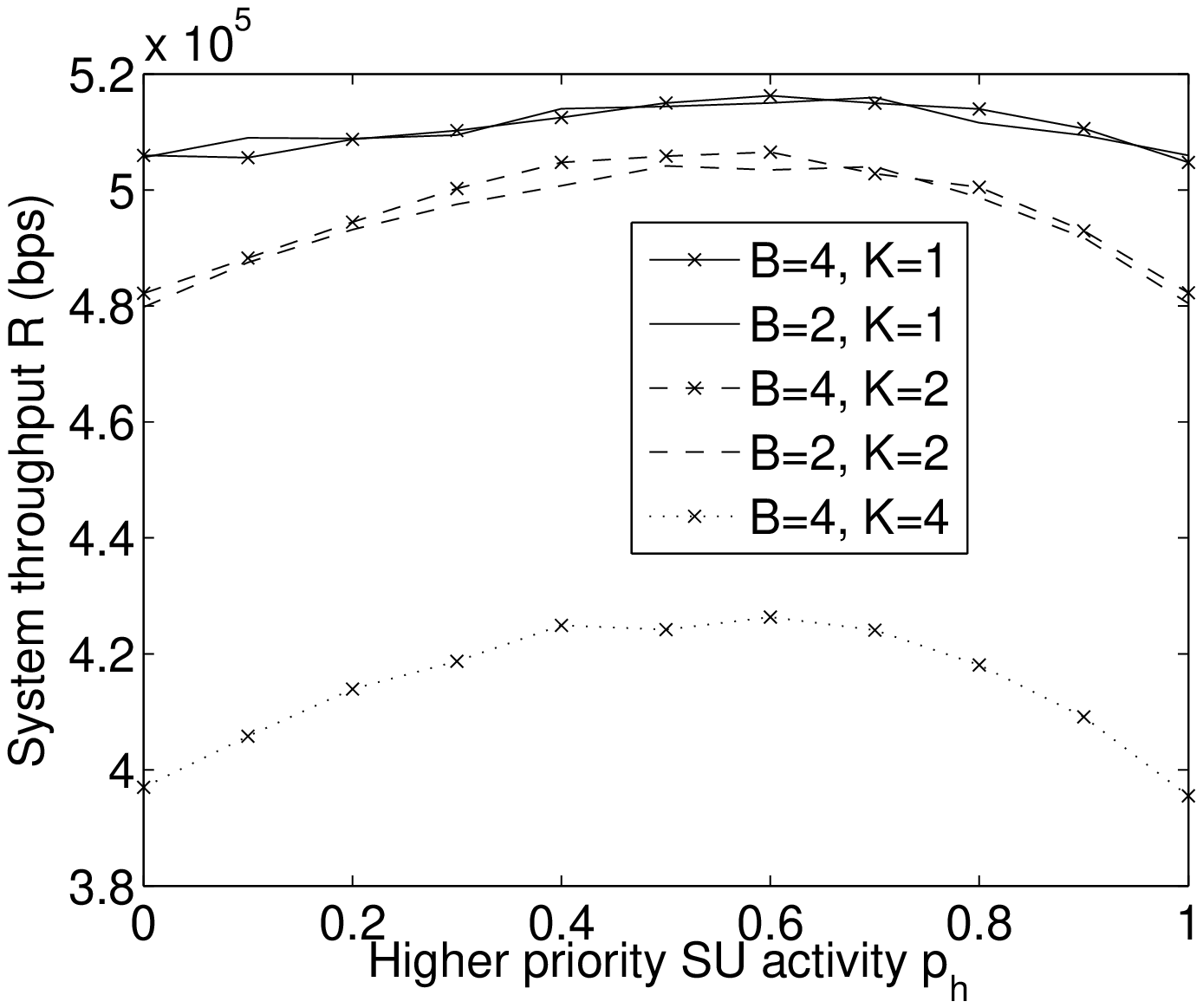}\label{fig:ph_thr_small}}
\subfigure[]{\includegraphics[width=0.49\columnwidth]{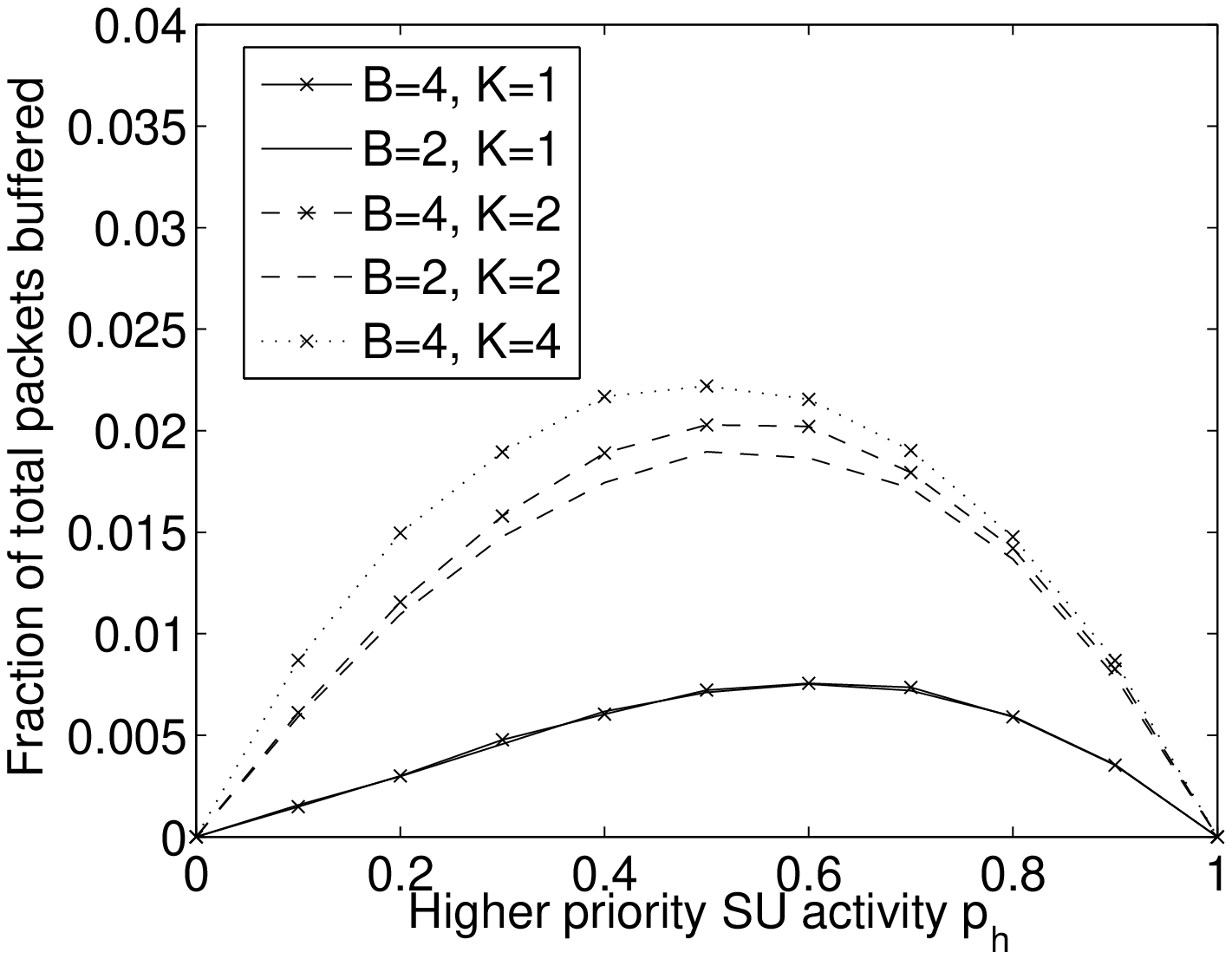}\label{fig:ph_prm_small}}
\caption{Fairness of prioritization in flexible multichannel MAC as a function of prioritization rate $p_h$: (a) throughput, (b) low priority SU packet buffering rate; for three bonding orders, $K\in\{1,2,4\}$ and two different buffer sizes, $B\in\{2,4\}$. Small network size is considered, $q_p=0.05$ while the rest of the common parameters are the same as listed for Fig.~\ref{fig:impact_qp}.}
\label{fig:impact_priority}
\end{figure}

We consider fairness by observing simultaneously, for one (small scale) network setup\footnote{Note that we do not present accompanying results for large scale OSA network, as the buffering happens seldom in this case becoming statistically negligible.}: (i) throughput obtained for the complete OSA network and (ii) average number of low priority SU frames (as a fraction of total SU frames) that are displaced to the buffer, both as a function of $p_h$. The results are presented in Fig.~\ref{fig:impact_priority} for three bonding orders, $K\in\{1,2,4\}$ and two different buffer sizes, $B\in\{2,4\}$. Observing the first metric, we see that the system throughput has a hyperbolic shape, maximized around $p_h=0.5$ and equal to system throughput with $B=0$ at $p_h\in\{0,1\}$, since it is at these two points that buffering is not effective. That is, when there are no high-priority connections ($p_h=0$) or when every connection is high-priority and cannot be displaced ($p_h=1$). We observe that the buffer size $B$ does not help in obtaining higher throughput for given value of $K$. Furthermore, an increase in $K$ results in a decline in throughput due to a decrease in the bond adaptivity of each generated connection.

The second metric, i.e. the fraction of total packets buffered presented in Fig.~\ref{fig:ph_prm_small}, is an important indicator of performance trade-off for the system. As the fraction of total buffered connections ($B$) increases, more connection disruptions occur for regular SU connections. Furthermore, we observe a trade-off with respect to $p_h$ since, for each curve, the highest fraction of buffered packets occurs approximately in the range of $p_h\in\{0.4,0.6\}$, causing SUs to occur maximum delay in reconnection. When observing frame buffering rate as a function of $p_h$, we see that the considered MAC protocol is the most fair for $p_h\in\{0,1\}$ demonstrating that the overall increase in system throughput incurs a penalty to regular connections. The shape of the packet buffering function is different for different combinations of $B$ and $K$. That is, the smaller the $K$, the more skewed towards higher $p_h$ the function is. With higher $K$ the effect of increasing $B$ and its related increase in fraction of packet buffers becomes more profound (conversely, observe that curves for `$B=4$, $K=1$' and `$B=2$, $K=1$' completely overlap).

The main message is to keep $B<M$ for the same value of $K$, as the proposed MAC protocol obtains the same throughout for the same $K$ irrespective of $B$ (see again Fig.~\ref{fig:ph_thr_small}), while the packet buffering rate increases as $B$ tends to $M$. Mean waiting times for the buffered connections are approximately equal for each case considered, specifically: (i) 4.59 slots for $B=4$, $K=1$, (ii) 4.55 slots for $B=2$, $K=1$, (ii) 4.73 slots for $B=4$, $K=2$, (iii) 4.49 slots for $B=2$, $K=2$, and (iv) 4.65 slots for $B=4$, $K=4$.

\subsection{Impact of Traffic/Network Dependent Adaptive Channel Bonding on System Throughput}
\label{sec:results_adaptive_bonding}

In the following study we extend the operation of the principal (flexible) channel bonding multichannel MAC protocol and $K$-only channel bonding multichannel MAC protocol with specialized channel adaptation features. In other words, the bonding order $K$ is adapted in an effort to maximize SU throughput. Such an observation can be made while observing Fig.~\ref{fig:impact_MN}, where depending on the user pool size, $N$, certain bonding orders obtain higher throughput than others.

Let us assume that within a time interval $i$ PU/SU traffic parameters are stationary. Then we can define the following optimization function for time interval $i$, necessary to find the the optimal bonding scheme:
\begin{subequations}\begin{eqnarray}
& \displaystyle\operatorname*{arg\,max}_{\mathbf{K}} R\label{eq:optimization1}\\
& \text{given } M,N, d, p, q_p, \label{eq:optimization2}\\
& \text{such that } K\leq K_{\max},p_{d}\geq p_{d,r},p_{f}\leq p_{f,r}, \label{eq:optimization3}
\end{eqnarray}
\end{subequations}
where $K=\kappa_i$ refer to the optimal bonding order at a specific value of $q_p$ for time interval $i$, $K_{\max}$ represents the maximum bonding level allowed, while $p_{f,r}$ and $p_{d,r}$ denote the OSA network-wide required probability of false alarm and mis-detection, respectively. Note that $R$ is a function of all parameters considered in (\ref{eq:optimization2})--(\ref{eq:optimization3}), while the framework to derive $R$ is presented in Section~\ref{sec:analytical_model}.

To explore the effectiveness of the proposed optimization framework, we consider two network scenarios identical to those considered in Section~\ref{sec:result_pu_impact}, i.e., a small and a large network scenario, with $K_{\max}=3$, $p_{d,r}=0.9$ and $p_{f,r}=0.1$. All other parameters remain the same as in Section~\ref{sec:result_pu_impact} except that PUs change their activity level in the course of the simulation from $q_p=0$ in increments of $\Delta=0.05$ to $q_p=0.1$. A small range of $q_p$ values were chosen in accordance with Fig.~\ref{fig:impact_qp} where it is observed that higher bond orders have better performance at $q_p \lessapprox 0.1$.

Furthermore, as a bonding-dependent parameter we consider PU activity. During each of the above-mentioned time intervals the bonding order is assumed to remain constant. We assume that $\zeta$ denotes the number of time intervals considered, while each $\kappa_i$ in the bonding scheme $\{\kappa_1,\cdots,\kappa_{\zeta}\}$ is the bonding order for an associated value of $q_p$. Note that we do not adapt $K$ on SU traffic conditions as they are more difficult to estimate in real-life network operation than PU statistics\footnote{For a recent discussion of efficient PU traffic estimation we refer e.g. to~\cite{gabran_jsac_subm} and the references therein.}. For example, length of SU frames are not known a priori to the receiver (such information is not sent in the RTS frame in our system model), thus other nodes cannot estimate for how long a virtual channel will be occupied by a certain sender/receiver pair just by listening to a DCC.

\begin{figure}
\centering
\subfigure[$M=4$, $N=12$]{\includegraphics[width=0.49\columnwidth]{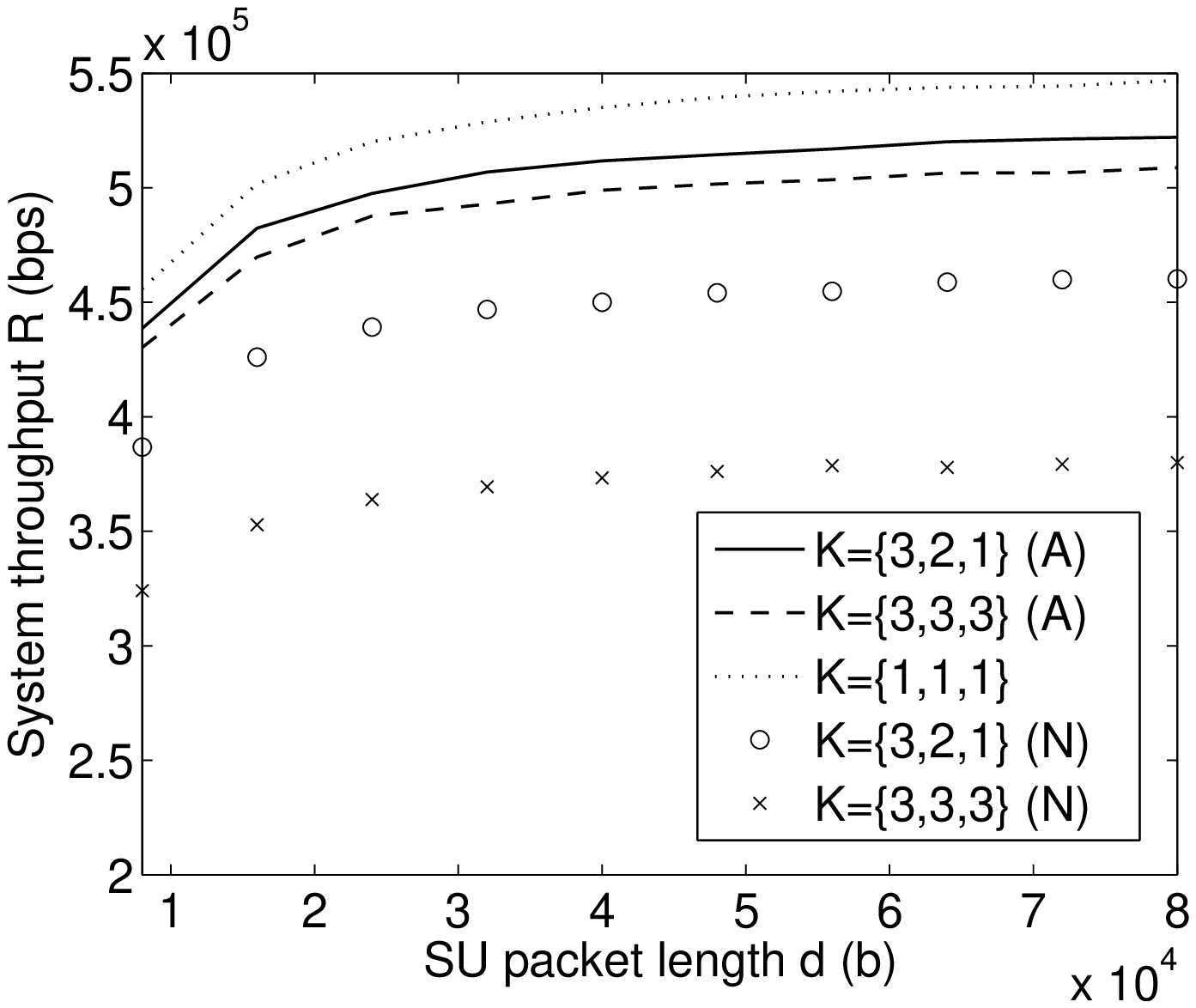}\label{fig:adapt_a}}
\subfigure[$M=12$, $N=40$]{\includegraphics[width=0.49\columnwidth]{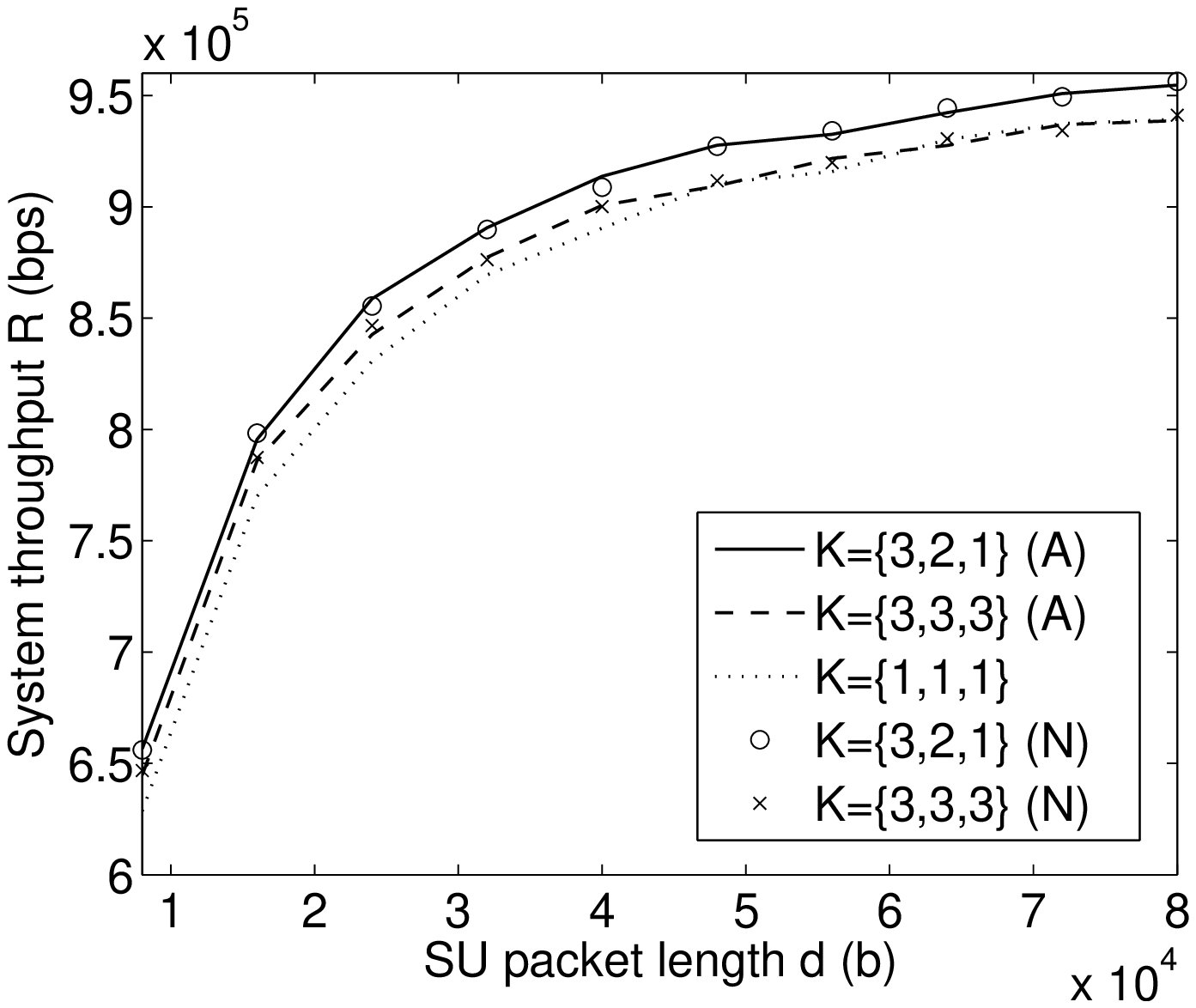}\label{fig:adapt_b}}
\caption{Impact of an adaptive channel bonding scheme on system throughput for (a) small network and (b) large network for three adaptive channel bonding schemes, i.e. $K=\{3,2,1\}$, $K=\{3,3,3\}$ and $K=\{1,1,1\}$ and two channel bonding MAC protocols, i.e. flexible (A) and $K$-only (N), as a function of the SU frame length $d$ for PU activity of $q_p\in\{0,0.05,0.1\}$; set of common parameters is the same as in Fig.~\ref{fig:impact_qp}.}
\label{fig:impact_adaptive}
\end{figure}

In Fig~\ref{fig:adapt_a} we present the SU throughput performance for $\zeta=3$ and three different bonding schemes for a small network scenario and corresponding $q_p$ values: an optimal bonding scheme of $\{1,1,1\}$, obtained via solving (\ref{eq:optimization1})--(\ref{eq:optimization2}), and for comparison $\{3,2,1\}$ and $\{3,3,3\}$. We immediately notice that bonding, i.e. $K>1$ for any value of $q_p$, does not provide any gains and is in fact more harmful in terms of SU throughput. This is attributed to the bonding order being nearly equal to the channel capacity of the system, $M$, therefore many incoming SU connections that are granted access through the control channel are blocked. Furthermore, the collision resolution strategy used in the system makes it unfavorable for higher bond orders to maintain a connection because if a PU happens to occupy any single physical channel in a bonded virtual channel the entire frame is preempted, as described in Section~\ref{sec:primary_user_detection}.

In Fig~\ref{fig:adapt_b} we represent the SU throughput performance in a large network scenario, for the respective $q_p$ values, considering optimal bonding scheme of $\{3,2,1\}$, again obtained via solving (\ref{eq:optimization1})--(\ref{eq:optimization2}), and for comparison $\{3,3,3\}$ and $\{1,1,1\}$ bonding schemes. In the large network scenario $M>K$, therefore a few bonded connections do not occupy the entire system bandwidth making it possible for multiple bonded connections to be present. This phenomenon allows the adaptive bonding scheme to utilize higher bond orders in time intervals in which $q_p \approxeq 0$ in order to maximize the usage of channels. This is clearly visible as the adaptive bonding schemes $\{3,2,1\}$ and $\{3,3,3\}$ outperform the single-bonded scheme $\{1,1,1\}$. Furthermore, $\{3,2,1\}$ outperforms $\{3,3,3\}$ as it is able to reduce the bonding order with an increase in PU activity level.

In addition, as concluded in Section~\ref{sec:result_pu_impact}, the $K$-only bonding scheme performs worse than the flexible channel bonding scheme for small-scale networks. This is because of the possibility of using a fixed number of channels, i.e. $K$ in this particular case, for each connection is smaller then that of a large-scale network (notice the large difference in performance for the two protocol types in Fig.~\ref{fig:adapt_a}). However, as the system size increases, the difference between the flexible and the $K$-only channel bonding MAC diminishes because there is a higher probability $K$-order connections are made due to increased contentions on the control channel. In the case considered in this paper, the difference between the two protocol options is almost negligible, see Fig.~\ref{fig:adapt_b}.

\subsection{Impact of Heavy-Tailed Distribution of PU Traffic on the Performance of Flexible Channel Bonding Protocol}
\label{sec:non_geometric}

So far we have assumed that the PU traffic is distributed according to the geometric distribution. In this section we relax this assumption by considering the impact of a heavy-tailed distribution on the channel bonding MAC protocol performance, in the same way as it was considered in~\cite[Sec. 5.2.6]{park_tmc_2011} for the non-bonding multichannel MAC protocols. In particular, we assess how the long-tail distribution of the PU activity affects the performance of the flexible channel bonding MAC protocol. As an example of a long-tail distribution we use the log-normal distribution, as it was concluded in~\cite[Table 3 and 4]{wellens_phycom_2009} that measured PU occupancies can be described by such a distribution in approximately 40\% of the cases for GSM 900 uplink, GSM 1800 downlink, DECT and 2.4\,GHz UNII channels.

\begin{figure}
\centering
\subfigure[]{\includegraphics[width=0.49\columnwidth]{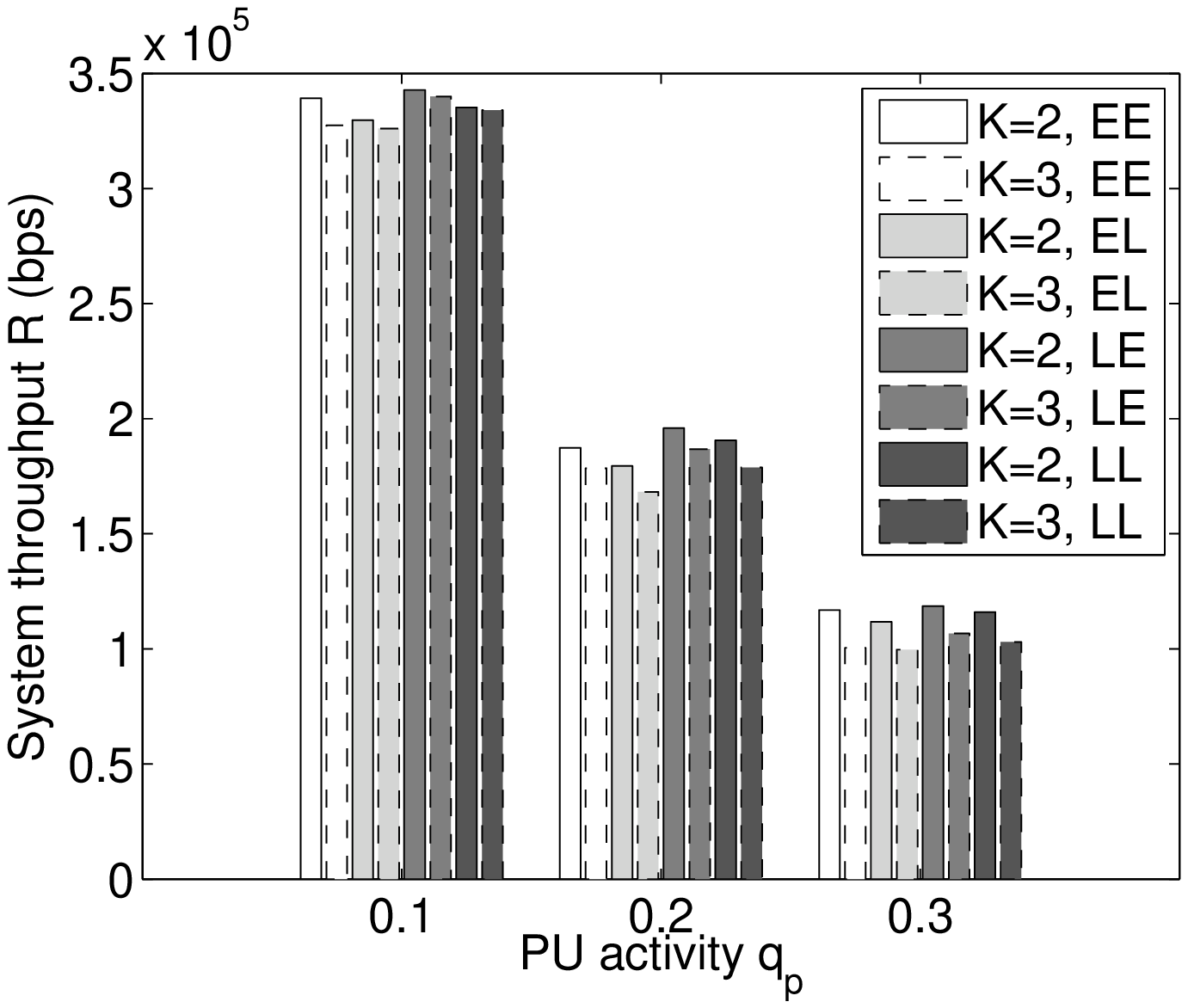}\label{fig:logn_M4N12}}
\subfigure[]{\includegraphics[width=0.49\columnwidth]{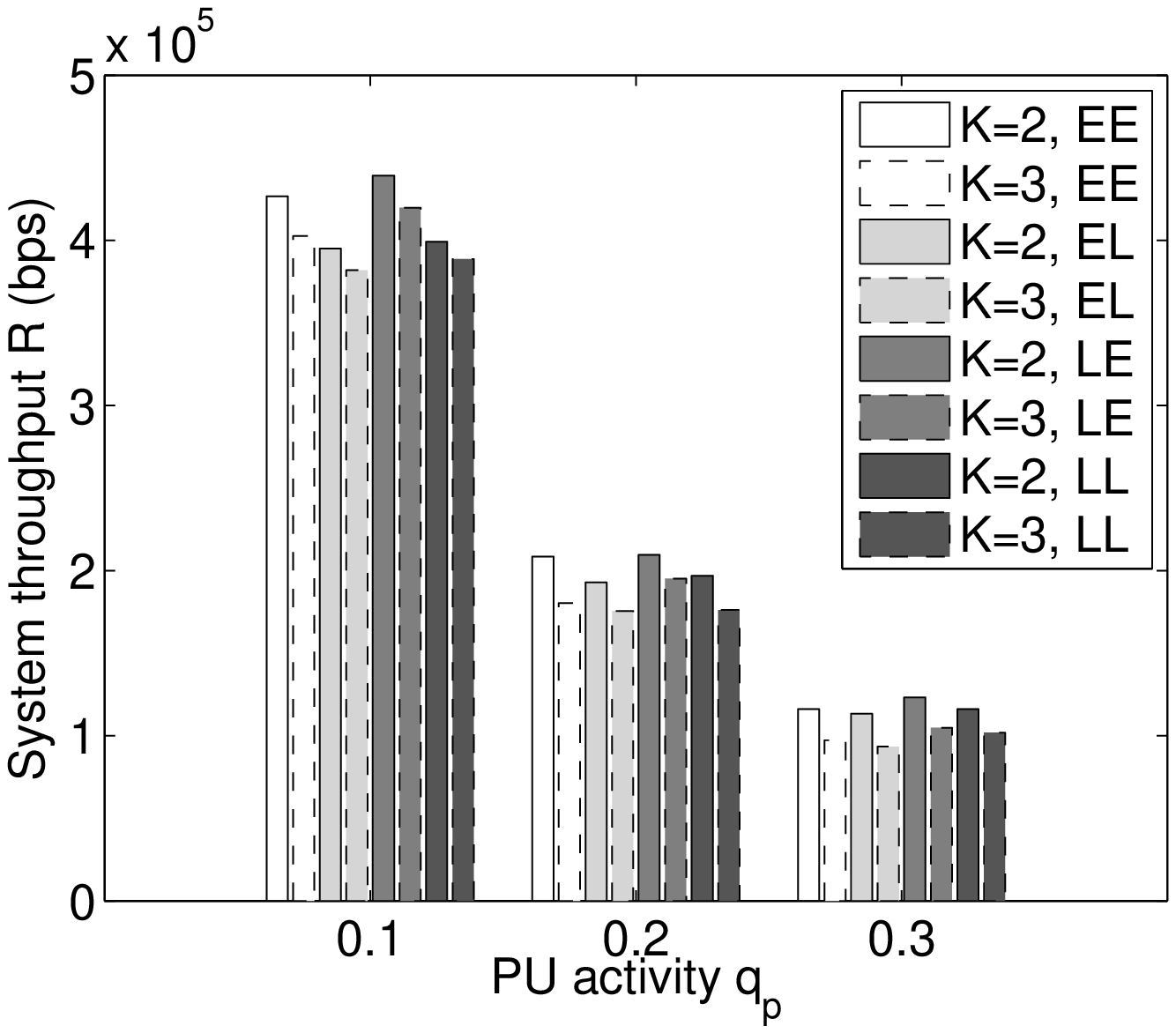}\label{fig:logn_M12N40}}
\caption{Impact of different PU ``on'' and ``off'' times distributions on OSA flexible channel bonding MAC protocol performance: (a) small network, and (b) large network, as given in Fig.~\ref{fig:impact_qp}. Legend symbols `E', and `L' denote geometric and log-normal distribution respectively, while the first and the second parameter of the two-letter symbol denote ``off'' and ``on'' time, respectively.}
\label{fig:lognormal_distribution_impact}
\end{figure}

We present the protocol performance for four different combinations of ``on'' and ``off'' time distributions, considering geometric and log-normal distribution, i.e. all possible combinations of ``on'' and ``off'' times obtained in~\cite[Table 3 and 4]{wellens_phycom_2009}. They are denoted symbolically as: (i) EE, (ii) EL, (iii) LE and (iv) LL, where `E' and `L' denote the geometric and the log-normal distribution, respectively, while the first and second position within the two-letter expression denote ``off'' and ``on'' times, respectively. Just like in~\cite[Sec. 5.2.6]{park_tmc_2011}, the parameter of the log-normal distribution was selected such that its mean value was equal to $q_l=1/q_p$  for the ``on'' time and $q_l=1-1/q_p$ for the ``off'' time. Because the log-normal distribution is continuous, it was rounded to the nearest integer, with the scale parameter $\sigma=\sqrt{\log\left({v_{l}}/{c_{l}^2}+1\right)}$ and location parameter $\mu=\log\left({c_{l}^2}{\sqrt{v_{l}+c_{l}^2}}^{-1}\right)$, where $c_l=1/q_{l}$, $v_l=(1-q_{l})/q_{l}^2$ is the mean and variance of the resulting discretized log-normal distribution. Note, as in~\cite[Sec. 5.2.6]{park_tmc_2011}, that the variance of the resulting discretized log-normal distribution is equal to the variance of the geometric distribution for the same mean value.

The results are presented for two types of networks denoted symbolically as small and large (see Fig~\ref{fig:impact_qp} for details), respectively in Fig~\ref{fig:logn_M4N12} and Fig.~\ref{fig:logn_M12N40}. The values of PU activity, $q_p\in\{0.1, 0.2, 0.3\}$, represent low activity rates, which are of the most interest for the network designer. Furthermore, for the sake of clarity of the presentation we consider only two bonding orders, i.e. $K\in\{2,3\}$. First, we observe that the impact of log-normal PU channel occupancy distribution on the system throughput is not large as one might expect (for example for $q_p=0.3$ and the large network, Fig.~\ref{fig:logn_M12N40}, the difference between the EE and LE cases is less than 7\%). If one wants to select the best combination of PU distribution that maximizes system throughput, it would be LE followed by LL. The impact of the non-geometric distribution on throughput is more visible for large networks than for smaller ones, compare Fig~\ref{fig:logn_M4N12} with Fig.~\ref{fig:logn_M12N40} as in large networks there are more opportunities to exploit larger number of channels for transmission. All bonding orders are effected equally by the non-geometric process of the PU occupancy, thus all conclusions drawn for the channel bonding MAC protocol using the geometrically distributed PU traffic will also hold for other combinations of PU traffic.

\section{Conclusions }
\label{sec:conclusions}

In this paper we have developed a detailed analytical model for assessing system throughput of an ad-hoc Opportunistic Spectrum Access (OSA) network with channel bonding. Our analysis has led to a set of important conclusions. Firstly, we note that there is generally a benefit of channel bonding for OSA networks which is most prominent for low Primary User (PU) channel activity and/or large PU channel pools. On the other hand, in certain cases, channel bonding might result in significant throughput loss in comparison to a classical non-bonded system, i.e. when the PU activity is large or the users per channel ratio is very large. Secondly, to be able to exploit channel throughput fully, a medium access control protocol should adaptively change the channel bond order $K$ depending on the network conditions (in particular the number of available resources and currently contending users). Thirdly, with certain physical layer constraints on maximum channel capacity, e.g. due to the limited transmission power per channel, irrespective of its bandwidth, the benefit of channel bonding for ad-hoc networks is easily lost. However, networks with a small users per channel ratio are still able to obtain higher system throughput than a regular non-bonded system, even when these constraints are present, provided, again, that the PU activity level is small.



\begin{thebibliography}{10}
\providecommand{\url}[1]{#1}
\csname url@samestyle\endcsname
\providecommand{\newblock}{\relax}
\providecommand{\bibinfo}[2]{#2}
\providecommand{\BIBentrySTDinterwordspacing}{\spaceskip=0pt\relax}
\providecommand{\BIBentryALTinterwordstretchfactor}{4}
\providecommand{\BIBentryALTinterwordspacing}{\spaceskip=\fontdimen2\font plus
\BIBentryALTinterwordstretchfactor\fontdimen3\font minus
  \fontdimen4\font\relax}
\providecommand{\BIBforeignlanguage}[2]{{%
\expandafter\ifx\csname l@#1\endcsname\relax
\else
\language=\csname l@#1\endcsname
\fi
#2}}
\providecommand{\BIBdecl}{\relax}
\BIBdecl

\bibitem{joshi_globecom_2012}
S.~{Joshi}, P.~{Pawe{\l}czak}, D.~{Cabric}, and J.~{Villasenor}, ``Performance
  of channel bonding for opportunistic spectrum access networks,'' in
  \emph{Proc. IEEE GLOBECOM}, Anaheim, CA, USA, Dec. 3--7, 2012.

\bibitem{chandra_sigcomm_2008}
R.~{Chandra}, R.~{Mahajan}, T.~{Moscibroda}, R.~{Raghavendra}, and P.~{Bahl},
  ``A case for adapting channel width in wireless networks,'' in \emph{Proc.
  ACM SIGCOMM}, Seattle, WA, USA, Aug. 17--22, 2008.

\bibitem{arslan_conext_2010}
M.~Y. {Arslan}, K.~{Pelechrinis}, I.~{Broustis}, S.~V. {Krishnamurthy},
  S.~{Adepalli}, and K.~{Papagiannaki}, ``Auto-configuration of {IEEE} 802.11n
  {WLANs},'' in \emph{Proc. ACM CoNEXT}, Philadelphia, PA, USA, Nov. 30~--~Dec.
  3, 2010.

\bibitem{tan_sigcomm_2010}
K.~{Tan}, J.~{Fang}, Y.~{Zhang}, S.~{Chen}, L.~{Shi}, J.~{Zhang}, and
  Y.~{Zhang}, ``Fine-grained channel access in wireless {LAN},'' in \emph{Proc.
  ACM SIGCOMM}, New Dehli, India, Aug. 30~--~Sep. 3, 2010.

\bibitem{moscibroda_icnp_2008}
T.~{Moscibroda}, R.~{Chandra}, Y.~{Wu}, S.~{Sengupta}, P.~{Bahl}, and
  Y.~{Yuan}, ``Load-aware spectrum distribution in wireless {LANs},'' in
  \emph{Proc. IEEE ICNP}, Orlando, FL, USA, Oct. 19--22, 2008.

\bibitem{texas_wp_2003}
``{WLAN} channel bonding: Causing greater problems than it solves,'' Texas
  Instruments, Tech. Rep. SPLY003, Sep. 2003.

\bibitem{park_tmc_2011}
J.~{Park}, P.~{Pawe{\l}czak}, and D.~{\v{C}abri\'{c}}, ``Performance of joint
  spectrum sensing and {MAC} algorithms for multichannel opportunistic spectrum
  access ad hoc networks,'' \emph{{IEEE} Trans. Mobile Comput.}, vol.~10,
  no.~7, pp. 1011--1027, Jul. 2011.

\bibitem{mo_tmc_2008}
J.~{Mo}, H.-S.~W. {So}, and J.~{Walrand}, ``Comparison of multichannel {MAC}
  protocols,'' \emph{{IEEE} Trans. Mobile Comput.}, vol.~7, no.~1, pp. 50--65,
  Jan. 2008.

\bibitem{cao_dyspan_2010}
L.~{Cao}, L.~{Yang}, and H.~{Zheng}, ``The impact of frequency agility on
  dynamic spectrum sharing,'' in \emph{Proc. IEEE DySPAN}, Singapore, Apr.
  6--10, 2010.

\bibitem{yuan_mobihoc_2007}
Y.~{Yuan}, P.~{Bahl}, R.~{Chandra}, T.~{Moscibroda}, and Y.~{Wu}, ``Allocating
  dynamic time-spectrum blocks for cognitive radio networks,'' in \emph{Proc.
  ACM MobiHoc}, Montreal, Quebec, Canada, Sep. 9--14, 2007.

\bibitem{bianchi_jsac_2000}
G.~{Bianchi}, ``Performance analysis of of the {IEEE} 802.11 distributed
  coordination function,'' \emph{{IEEE} J. Sel. Areas Commun.}, vol.~18, no.~3,
  pp. 535--547, Mar. 2000.

\bibitem{bansal_ohiotech_2010}
T.~{Bansal}, D.~{Li}, and P.~{Sinh}, ``Fairness by sharing: Split channel
  allocation for cognitive radio networks,'' Ohio State University, Tech. Rep.
  Department of Computer Science and Engineering TR28, 2010.

\bibitem{Park_arxiv_2010}
J.~{Park}, P.~Pawe{\l}czak, P.~{Gr{\o}nsund}, and D.~\v{C}abri\'{c}, ``Analysis
  framework for opportunistic spectrum {OFDMA} and its application to the
  {IEEE} 802.22 standard,'' \emph{{IEEE} Trans. Veh. Technol.}, vol.~61, no.~5,
  pp. 2271--2293, Jun. 2012.

\bibitem{bahl_sigcomm_2009}
P.~{Bahl}, R.~{Chandra}, T.~{Moscibroda}, R.~{Murty}, and M.~{Welsh}, ``White
  space networking with {Wi-Fi} like connectivity,'' in \emph{Proc. ACM
  SIGCOMM}, Barcelona, Spain, EU, Aug. 17--21, 2009.

\bibitem{yuan_dyspan_2007}
Y.~{Yuan}, P.~{Bahl}, R.~{Chandra}, and P.~A. {Chou}, ``{KNOWS}: Cognitive
  radio networks over white spaces,'' in \emph{Proc. IEEE DySPAN}, Dublin,
  Ireland, EU, Apr. 17--20, 2007.

\bibitem{yang_nsdi_2010}
L.~{Yang}, W.~{Hou}, L.~{Cao}, B.~Y. {Zhao}, and H.~{Zheng}, ``Supporting
  demanding wireless applications with frequency-agile radios,'' in \emph{Proc.
  NSDI}, San Jose, CA, USA, Apr. 28--30, 2010.

\bibitem{kone_imc_2010}
V.~{Kone}, L.~{Yang}, X.~{Yang}, B.~Y. {Zhao}, and H.~{Zheng}, ``On the
  feasibility of effective opportunistic spectrum access,'' in \emph{Proc. ACM
  IMC}, Melbourne, Australia, Nov. 1--3, 2010.

\bibitem{pawelczak_tvt_2009}
P.~Pawe{\l}czak, S.~{Pollin}, H.-S.~W. {So}, A.~{Bahai}, R.~V. {Prasad}, and
  R.~{Hekmat}, ``Performance analysis of multichannel medium access control
  algorithms for opportunistic spectrum access,'' \emph{{IEEE} Trans. Veh.
  Technol.}, vol.~58, no.~6, pp. 3014--3031, Jul. 2009.

\bibitem{jha_twc_2011}
S.~C. {Jha}, U.~{Phuyal}, M.~M. {Rashid}, and V.~K. {Bhargava}, ``Design of
  {OMC-MAC}: An opportunistic multi-channel {MAC} with {QoS} provisioning for
  distributed cognitive radio networks,'' \emph{{IEEE} Trans. Wireless
  Commun.}, vol.~10, no.~10, pp. 3414--3425, Oct. 2011.

\bibitem{baldo_twc_2010}
N.~{Baldo}, A.~{Asterjadhi}, and M.~{Zorzi}, ``Dynamic spectrum access using a
  network coded cognitive control channel,'' \emph{{IEEE} Trans. Wireless
  Commun.}, vol.~9, no.~8, pp. 2575--2587, Aug. 2010.

\bibitem{zhang_jstsp_2011}
X.~{Zhang} and H.~{Su}, ``{CREAM-MAC}: Cognitive radio-en{A}bled multi-channel
  mac protocol over dynamic spectrum access networks,'' \emph{{IEEE} J. Select.
  Topics Signal Processing}, vol.~5, no.~1, pp. 110--123, Feb. 2011.

\bibitem{lin_tvt_2012}
Y.-Y. {Lin} and K.-C. {Chen}, ``Asynchronous dynamic spectrum access,''
  \emph{{IEEE} Trans. Veh. Technol.}, vol.~61, no.~1, pp. 222--236, Jan. 2012.

\bibitem{wang_twc_2012}
S.~{Wang}, J.~{Zhang}, and L.~{Tong}, ``A characterization of delay performance
  of cognitive medium access,'' \emph{{IEEE} Trans. Wireless Commun.}, vol.~11,
  no.~22, pp. 800--809, Feb. 2012.

\bibitem{Liang_twc_2008}
Y.-C. {Liang}, Y.~{Zeng}, E.~C. {Peh}, and A.~T. {Hoang}, ``Sensing throughput
  tradeoff in cognitive radio networks,'' \emph{{IEEE} Trans. Wireless
  Commun.}, vol.~7, no.~4, pp. 1326--1337, Apr. 2008.

\bibitem{Hoang_twc_2009}
A.~T. {Hoang}, Y.-C. {Liang}, D.~T.~C. {Wong}, Y.~{Zeng}, and R.~{Zhang},
  ``Opportunistic spectrum access for energy-constrained cognitive radios,''
  \emph{{IEEE} Trans. Wireless Commun.}, vol.~8, no.~3, pp. 1206--1211, Mar.
  2009.

\bibitem{wangchun_jsac_2011}
L.-C. {Wang}, C.-W. {Wang}, and F.~{Adachi}, ``Load-balancing spectrum decision
  for cognitive radio networks,'' \emph{{IEEE} J. Sel. Areas Commun.}, vol.~29,
  no.~4, pp. 757--769, Apr. 2011.

\bibitem{cheng_jsac_2011}
H.~T. {Cheng} and W.~{Zhuang}, ``Simple channel sensing order in cognitive
  radio networks,'' \emph{{IEEE} J. Sel. Areas Commun.}, vol.~29, no.~4, pp.
  676--688, Apr. 2011.

\bibitem{bian_jsac_2011}
K.~{Bian}, J.-M. {Park}, and R.~{Chen}, ``Control channel establishment in
  cognitive radio networks using channel hopping,'' \emph{{IEEE} J. Sel. Areas
  Commun.}, vol.~29, no.~4, pp. 689--703, Apr. 2011.

\bibitem{Papadimitratos_commag_2005}
P.~{Papadimitratos}, S.~{Sankaranarayanan}, and A.~{Mishra}, ``A bandwidth
  sharing approach to improve licensed spectrum utilization,'' \emph{{IEEE}
  Commun. Mag.}, vol.~43, no.~12, pp. S10--S14, Dec. 2005.

\bibitem{gronsund_pimrc_2009}
P.~{Gr{\o}nsund}, H.~N. {Pham}, and P.~E. {Engelstad}, ``Towards dynamic
  spectrum access in primary {OFDMA} systems,'' in \emph{Proc. IEEE PIMRC},
  Tokyo, Japan, Sep. 13--16, 2009.

\bibitem{gambini_twc_2008}
J.~{Gambini}, O.~{Simeone}, Y.~{Bar-Ness}, U.~{Spagnolini}, and T.~{Yu},
  ``Packet-wise vertical handover for unlicensed multi-standard spectrum access
  with cognitive radios,'' \emph{{IEEE} Trans. Wireless Commun.}, vol.~7,
  no.~12, pp. 5172--5176, Dec. 2008.

\bibitem{Gerihofer_commag_2007}
S.~{Geirhofer}, L.~{Tong}, and B.~M. {Sandler}, ``Dynamic spectrum access in
  the time domain: Modeling and exploiting white space,'' \emph{{IEEE} Commun.
  Mag.}, vol.~45, no.~5, pp. 66--72, May 2007.

\bibitem{Huang_infocom_2009}
S.~{Huang}, X.~{Liu}, and Z.~{Ding}, ``Optimal sensing-transmission structure
  for dynamic spectrum access,'' in \emph{Proc. IEEE INFOCOM}, Rio De Janeiro,
  Brazil, Apr. 19--25, 2009.

\bibitem{deng_tcom_2002}
Z.~J. {Haas} and J.~{Deng}, ``Dual busy tone multiple access ({DBTMA})---a
  multiple access control scheme for ad hoc network,'' \emph{{IEEE} Trans.
  Commun.}, vol.~50, no.~6, pp. 975--985, Jun. 2002.

\bibitem{salameh_arizonatech_2011}
H.~B. {Salameh}, M.~{Krunz}, and D.~{Manzi}, ``Design and evaluation of an
  efficient guard-band-aware multi-channel spectrum sharing mechanism,''
  University of Arizona, Tucson, AZ, USA, Tech. Rep. TR-UA-ECE-2011-1, 2011.

\bibitem{geirhofer_commag_2007}
S.~{Geirhofer}, L.~{Tong}, and B.~M. {Sadler}, ``Dynamic spectrum access in the
  time domain: Modeling and exploiting white spaces,'' \emph{{IEEE} Commun.
  Mag.}, vol.~45, no.~5, pp. 66--72, May 2007.

\bibitem{wellens_phycom_2009}
M.~{Wellens}, J.~{Riihij{\"a}rvi}, and P.~{M{\"a}h{\"o}nen}, ``Empirical time
  and frequency domain models of spectrum use,'' \emph{Elsevier Physical
  Communication Journal}, vol.~2, no. 1--2, pp. 10--32, Mar.--Jun. 2009.

\bibitem{wang_jsac_2011}
X.~Y. {Wang}, A.~{Wong}, and P.-H. {Ho}, ``Stochastic medium access for
  cognitive radio ad hoc network,'' \emph{{IEEE} J. Sel. Areas Commun.},
  vol.~29, no.~4, pp. 770--783, Apr. 2011.

\bibitem{wang_tvt_2010}
P.~{Wang}, D.~{Niyato}, and H.~{Jiang}, ``Voice-service capacity analysis for
  cognitive radio networks,'' \emph{{IEEE} Trans. Veh. Technol.}, vol.~59,
  no.~4, pp. 1779--1790, May 2010.

\bibitem{zhou_ieeetwc_2008}
P.~{Zhou}, H.~{Hu}, H.~{Wang}, and H.-H. {Chen}, ``An efficient random access
  scheme for {OFDMA} systems with implicit message transmission,'' \emph{{IEEE}
  Trans. Wireless Commun.}, vol.~7, no.~7, pp. 2790--2797, Jul. 2008.

\bibitem{hassel_twc_2007}
V.~H. G.~E. {{\O}ien} and D.~{Gesber}, ``Throughput guarantees for wireless
  networks with opportunistic scheduling: A comparative study,'' \emph{{IEEE}
  Trans. Wireless Commun.}, vol.~6, no.~12, pp. 4215--4220, Dec. 2007.

\bibitem{jain_tr_1984}
R.~K. {Jain}, D.-M.~W. {Chiu}, and W.~R. {Have}, ``A quantitative measure of
  fairness and discrimination for resource allocation in shared computer
  systems,'' Digital Equipment Corporation, Hudson, MA, Tech. Rep. DEC-TR-301,
  Sep. 1984.

\bibitem{gabran_jsac_subm}
\BIBentryALTinterwordspacing
W.~{Gabran}, C.-H. {Liu}, P.~{Pawe{\l}czak}, and D.~{\v{C}abri\'{c}}. (2012)
  Primary user traffic estimation for dynamic spectrum access. [Online].
  Available: \url{http://arxiv.org/abs/1203.2393}
\BIBentrySTDinterwordspacing

\end{thebibliography}
\end{document}